\documentclass[aps,10pt,preprintnumbers,showkeys,eqsecnum,%
superscriptaddress,nofootinbib,showpacs]{revtex4-1}
\usepackage{bm} 
\usepackage{graphicx}
\usepackage{hyperref}
\newcommand{\be}{\begin{eqnarray}}
\newcommand{\ee}{\end{eqnarray}}
\newcommand{\beq}{\begin{equation}}
\newcommand{\eeq}{\end{equation}}
\begin{document}
\title{Light--front representation of chiral dynamics in peripheral 
transverse densities}
\author{C.~Granados}
\email[E--mail: ]{carlos.granados@physics.uu.se}
\affiliation{Department of Physics and Astronomy, 
Uppsala University, Box 516, 75120 Uppsala, Sweden}
\author{C.~Weiss}
\email[E--mail: ]{weiss@jlab.org}
\affiliation{Theory Center, Jefferson Lab, Newport News, VA 23606, USA}
\begin{abstract}
The nucleon's electromagnetic form factors are expressed in 
terms of the transverse densities of charge and magnetization at 
fixed light--front time. At peripheral transverse distances 
$b = O(M_\pi^{-1})$ the densities are governed by chiral dynamics 
and can be calculated model--independently using chiral effective field 
theory (EFT). We represent the leading--order chiral EFT results for
the peripheral transverse densities as overlap integrals of chiral
light--front wave functions, describing the transition of the initial 
nucleon to soft pion--nucleon intermediate states and back. 
The new representation (a)~explains the parametric order of the 
peripheral transverse densities; (b)~establishes an inequality between 
the spin--independent and --dependent densities; 
(c)~exposes the role of pion orbital angular momentum in chiral dynamics; 
(d)~reveals a large left--right asymmetry of the current in a 
transversely polarized nucleon and suggests a simple interpretation.
The light--front representation enables a first--quantized, 
quantum--mechanical view of chiral dynamics that is fully relativistic 
and exactly equivalent to the second--quantized, field--theoretical 
formulation. It relates the charge and magnetization densities 
measured in low--energy elastic scattering to the generalized parton 
distributions probed in peripheral high--energy scattering processes. 
The method can be applied to nucleon form factors of other operators, 
e.g.\ the energy--momentum tensor.
\end{abstract}
\preprint{JLAB-THY-15-2018}
\maketitle
\tableofcontents
\newpage
\section{Introduction}
\label{sec:introduction}
Transverse densities have become an essential tool in the analysis of 
current matrix elements (vector, axial) and the description of the spatial 
structure of 
hadrons \cite{Soper:1976jc,Burkardt:2000za,Miller:2007uy,Miller:2010nz}. 
They are defined as the two--dimensional Fourier transforms of the 
invariant form factors and describe the distribution of charge 
and current in the hadron in transverse space at fixed light--front 
time $x^+ \equiv t + z$. 
They are frame--independent (boost--invariant) and provide an objective 
spatial representation of the hadron as a relativistic system.
In the context of QCD the transverse densities correspond to a projection
of the generalized parton distributions (GPDs) describing the transverse 
spatial distribution of quarks and 
antiquarks \cite{Burkardt:2000za,Burkardt:2002hr}; 
as such they connect the information gained from low--energy elastic 
scattering with the partonic content probed by high--momentum--transfer 
processes in high--energy scattering. In composite models of nucleon 
structure the transverse densities can be
expressed as proper densities of the light--front wave functions of the 
system and are therefore a natural ground for phenomenological analysis.
Considerable efforts have been devoted to extracting the transverse 
charge and magnetization densities in the nucleon from the available
electromagnetic (Dirac, Pauli) form factor data and 
studying their properties \cite{Miller:2007uy,Carlson:2007xd,Venkat:2010by}; 
see Ref.~\cite{Miller:2010nz} for a review.

Of particular interest are the densities in the nucleon's chiral periphery. 
At transverse distances $b = O(M_\pi^{-1})$, where the pion mass is 
regarded as parametrically small compared to the typical inverse 
hadronic size, the densities are governed by the universal dynamics 
resulting from 
the spontaneous breaking of chiral symmetry and can be computed
from first principles using the methods of chiral effective 
field theory (EFT). The isovector charge and magnetization 
densities, $\rho_{1}(b)$ and $\rho_{2}(b)$, arise from chiral 
processes in which the current couples to the nucleon through 
exchange of a two--pion system with momenta $O(M_\pi)$ relative to the
nucleon. A detailed investigation of the properties of the peripheral 
transverse densities in leading--order (LO) relativistic chiral EFT was 
performed in Ref.~\cite{Granados:2013moa}. The densities decay
exponentially with a range given by the mass of the exchanged 
system, $2 M_\pi$ (``Yukawa tail''); their overall strength and the 
underlying power--like behavior in $b$ are determined by coupling of 
the two--pion exchange to the nucleon and exhibit a rich structure.
It was found that the transverse charge density $\rho_1^V (b)$ and the 
modified magnetization density 
$\widetilde\rho_2^V (b) \equiv (\partial/\partial b) 
\rho_2^V(b)/(2 M_N)$ are of the same order in the chiral expansion,
obey an approximate inequality $\widetilde\rho_2^V (b) < \rho_1^V(b)$,
and are numerically very close at the distances of interest, 
$b \gtrsim 1\, M_\pi^{-1}$. These findings represent model--independent 
features of the nucleon's chiral periphery and call for a simple 
explanation.

In Ref.~\cite{Granados:2013moa} the peripheral densities were calculated
in a dispersive representation, where they are expressed as integrals 
of the imaginary parts (or spectral functions) of the invariant form 
factors along the cut in timelike region at $t > 4 M_\pi^2$. 
This formulation makes it possible to use the well--known chiral EFT 
results for the invariant form factors and their spectral functions for 
the calculation of the transverse densities. While it allows one to 
derive all properties of interest, it does not provide a mechanical
picture of the chiral processes as pions ``moving about'' the nucleon
in space and time. Such a picture could be obtained in a time--ordered
representation of chiral EFT, where one works with the concepts of 
instantaneous configurations, time evolution, and the wave function
of the system. Since the transverse densities are defined at fixed 
light--front time $x^+$ it is natural to adopt light--front 
quantization \cite{Dirac:1949cp,Leutwyler:1977vy,Brodsky:1997de} 
and follow the 
evolution of the relevant chiral processes in light--front time. 
This representation might explain our earlier findings and provide 
new insight into the structure of the peripheral transverse densities.

Studying the space--time evolution of chiral dynamics in light--front
quantization is interesting also for methodological reasons, unrelated 
to the specific questions posed by transverse densities. The typical 
momentum of soft pions in the nucleon rest frame is $k = O(M_\pi)$ [the 
velocity is $v = O(1)$], and the typical energy of configurations is 
$E = O(M_\pi)$. Chiral dynamics thus represents an essentially 
relativistic system, in which pions ``appear'' and ``disappear'' 
through quantum fluctuations and the number of particles is generally 
not conserved. In equal--time quantization the particle number
observed at an instant changes under Lorentz boost, so that the 
wave function is essentially frame--dependent and no meaningful 
particle--based description of the theory can be constructed.
In light--front quantization the particle number is invariant
under boosts, the wave function is frame--independent, and a 
natural particle--based description is obtained \cite{Brodsky:1997de}. 
It represents the only known formulation that permits a consistent
first--quantized particle--based description of chiral processes.
Such a representation could significantly advance our understanding
of chiral dynamics.

In this article we study the nucleon's transverse charge and magnetization
densities in the chiral periphery in a first--quantized particle--based 
representation of chiral dynamics based on light--front quantization. 
The LO chiral EFT results for the peripheral densities are 
expressed in time--ordered form, as the result of a transition of the
bare nucleon to a virtual $\pi N$ state mediated by the chiral EFT
interactions. The densities appear as overlap integrals of the perturbative
light--front wave functions describing the $N \rightarrow \pi N$
transition, which are calculable directly from the chiral Lagrangian.
The new representation offers new insight into the structure of
peripheral densities and reveals several interesting properties.
First, it explains in simple terms the parametric order of the peripheral 
charge and modified magnetization 
densities, $\rho_1^V (b)$ and $\widetilde\rho_2^V (b)$, 
in the chiral expansion. Second, it proves the inequality 
$|\widetilde\rho_2^V (b)| < \rho_1^V (b)$, which had been observed 
numerically in the earlier study using the invariant 
formulation \cite{Granados:2013moa}. 
It also explains why the inequality is approximately saturated, 
$\widetilde\rho_2^V (b) \approx \rho_1^V (b)$, and shows that this is
related to the essentially relativistic character of chiral dynamics.
Third, the wave function overlap representation exposes the role 
of the pion's orbital angular momentum in the peripheral transverse 
densities. A particularly simple picture is obtained with transversely 
polarized nucleon states, where only a single pion orbital with $L = 1$ 
accounts for both densities, and the relation between 
$\widetilde\rho_2^V (b)$ and $\rho_1^V (b)$ is explained by the 
``left--right'' asymmetry induced by the orbital motion of the pion 
in the preferred longitudinal direction. 
We emphasize that the first--quantized representation developed in the 
present work is \textit{equivalent} to the invariant LO chiral EFT 
expressions used in earlier studies, and that the new insights derived 
from it reflect general properties of the nucleon's chiral periphery.
A summary of our results has been presented in Ref.~\cite{Granados:2015lxa}.

In the present work we derive the light--front representation of 
peripheral chiral processes by rewriting the result obtained in 
Lorentz--invariant chiral EFT. This approach has the advantage that
it guarantees equivalence to the well--tested invariant formulation
and avoids the use of light--front specific techniques. The $\pi N$ 
light--front wave function of the nucleon in chiral EFT is defined in 
terms of the vertex function provided by the chiral Lagrangian, 
and the wave function overlap representation of the current matrix 
element is obtained naturally from the reduction of the Feynman integrals. 
The connection with the conventional light--front time--ordered 
Hamiltonian approach \cite{Brodsky:1997de}
is explained in Appendix~\ref{app:time-ordered}. Some formal
aspects of chiral EFT in the time--ordered formulation were studied 
in Refs.~\cite{Ji:2009jc,Burkardt:2012hk,Ji:2013bca}.
Our derivation also confirms the presence of an instantaneous term 
(or zero mode contribution) in the light--front representation of the 
chiral component of the transverse charge 
density \cite{Strikman:2010pu,Burkardt:2012hk}. 
This term has a simple physical interpretation as describing the 
contribution of large--mass (non--chiral) intermediate states in 
time--ordered chiral processes and is shown to be numerically small. 

The light--front representation of chiral dynamics described here can 
be applied also to the peripheral transverse densities of other 
local operators, e.g.\ the matter and momentum densities associated 
with the energy--momentum tensor, and to the peripheral GPDs probed 
in high--energy scattering processes. We derive the wave function 
overlap representation of the light--front ``plus'' momentum density 
of peripheral pions in the nucleon, which determine the chiral component 
of the nucleon's parton densities at transverse distances $b = O(M_\pi^{-1})$. 
This establishes a formal connection between our chiral EFT results for the 
peripheral transverse densities and the nucleon's quark/antiquark content. 
The light--front momentum density of peripheral pions could in principle
also be probed directly in peripheral high--energy scattering processes.

The plan of this article is as follows. In 
Sec.~\ref{sec:current_matrix_element} we review the transverse density
representation of the current matrix element, the peripheral chiral
contributions in the invariant formulation \cite{Granados:2013moa}, 
and derive the light--front overlap representation of the current 
matrix element. In Sec.~\ref{sec:wave_function} we investigate the
properties of the peripheral $\pi N$ light--front wave function,
including the choice of nucleon spin states and the coordinate 
representation. In Sec.~\ref{sec:densities} we express the transverse 
densities $\rho_1^V$ and $\widetilde\rho_2^V$ as overlap integrals of the 
coordinate--space $\pi N$ light--front wave functions and study 
their properties. Using longitudinal nucleon spin states we discuss the 
parametric order of the densities, derive the inequality between them,
and evaluate them numerically. We also present the expressions for
transverse nucleon spin states and show that they correspond to a simple
mechanical picture of a pion with $L = 1$ orbiting around the nucleon
in the rest frame. This picture concisely summarizes the dynamical
content of the LO chiral EFT contribution and represents
the main result of this work. We also compute the instantaneous
(contact term) contribution to the densities and show that it is
numerically small. In Sec.~\ref{sec:generalized} we connect the transverse 
densities with the peripheral parton content of the nucleon in QCD. 
We derive the wave function overlap representation of the pion 
plus momentum distribution (``pion GPD'') in chiral EFT, which 
determines the nucleon's peripheral parton 
densities \cite{Strikman:2003gz,Strikman:2009bd}, and show that the 
transverse charge density is recovered by integrating the peripheral 
quark/antiquark densities over the parton momentum fraction $x$.
\section{Transverse densities}
\label{subsec:transverse_densities}
The transition matrix element of the electromagnetic current between 
nucleon states is parametrized in terms of two invariant form factors
(we follow the notation and conventions of Ref.~\cite{Granados:2013moa})
\beq
\langle N(p_2, \sigma_2) | J^\mu (x) | N(p_1, \sigma_1) \rangle\
\;\; = \;\; \bar u_2 \left[ \gamma^\mu F_1(t) - 
\frac{\sigma^{\mu\nu} \Delta_\nu}{2 M_N} F_2(t) \right] u_1 \, 
e^{i\Delta x} ,
\label{me_general}
\eeq
where $p_{1, 2}$ are the nucleon 4--momenta, $\sigma_{1, 2}$ 
the spin quantum numbers, $u_1 \equiv u(p_1, \sigma_1)$ 
\textit{etc.}\ the nucleon bispinors, normalized to 
$\bar u_1 u_1 = \bar u_2 u_2 = 2 M_N$,
and $\sigma^{\mu\nu} \equiv \frac{1}{2}\left[ \gamma^\mu, \gamma^\nu\right]$.
The 4--momentum transfer is defined as 
\beq
\Delta \;\; \equiv \;\; p_2 - p_1,
\eeq
and the dependence of the matrix element on the space--time point $x$ 
where the current is measured is dictated by translational invariance.
The Lorentz--invariant momentum transfer is
\beq
t \;\; \equiv \;\; \Delta^2 \;\; = \;\; (p_2 - p_1)^2,
\eeq
with $t < 0$ in the physical region for electromagnetic scattering.
The Dirac and Pauli form factors, $F_1$ and $F_2$, are invariant
functions of $t$ and can be discussed independently of any reference frame. 

\begin{figure}[t]
\begin{center}
\includegraphics[width=.55\textwidth]{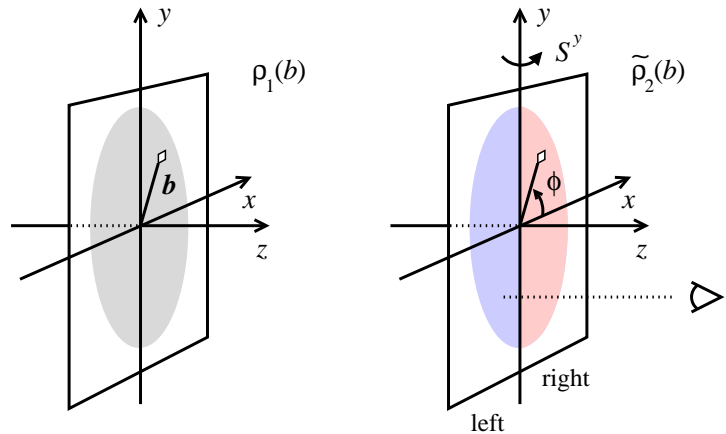}
\end{center}
\caption[]{Interpretation of the transverse densities of the 
electromagnetic current in a nucleon state with 
spin quantized in the transverse $y$--direction, Eq.~(\ref{j_plus_rho}). 
$\rho_1(b)$ describes the spin--independent (or left--right symmetric) 
part of the plus current density; $\cos\phi \, \widetilde\rho_2(b)$ 
describes the spin--dependent (or left--right asymmetric) part.}
\label{fig:interpretation}
\end{figure}

In the context of the light--front description of nucleon structure one 
naturally considers the form factors in a frame where the momentum 
transfer vector lies in the transverse ($x$--$y$) plane, 
\beq
\Delta^\mu \equiv (\Delta^0, \Delta^x, \Delta^y, \Delta^z) 
= (0, \bm{\Delta}_T, 0), \hspace{2em}
\bm{\Delta}_T = (\Delta^x, \Delta^y), \hspace{2em}
t = -\bm{\Delta}_T^2, 
\label{Delta_transverse}
\eeq
and represents them as Fourier transforms of certain spatial densities
\cite{Miller:2007uy,Miller:2010nz}
\beq
F_{1, 2}(t = -\bm{\Delta}_T^2) \;\; = \;\; \int d^2 b \; 
e^{i \bm{\Delta}_T \cdot \bm{b}} \; \rho_{1, 2} (b) ,
\label{rho_def}
\eeq
where $\bm{b} \equiv (b^x, b^y)$ is a transverse coordinate variable
and $b \equiv |\bm{b}|$. The formal properties of the transverse densities 
$\rho_{1, 2}(b)$ and their physical interpretation have been discussed
extensively in the literature \cite{Burkardt:2000za,Miller:2010nz}. 
and are summarized in Ref.~\cite{Granados:2013moa}. 
They describe the transverse spatial distribution of the light--front 
plus component of the current, $J^+ \equiv J^0 + J^z$, 
in the nucleon at fixed light--front time $x^+$. 
Specifically, in a state where the nucleon is localized in transverse 
space at the origin, and polarized in the $y$--direction, 
the matrix element of the current $J^+$ at light--front time
$x^+ = 0$ and light--front coordinates $x^- = 0$ and 
$\bm{x}_T = \bm{b}$ is given by
\begin{eqnarray}
\langle J^+ (\bm{b}) \rangle_{\text{\scriptsize localized}}
&=& (...) \; \left[
\rho_1 (b) \;\; + \;\; (2 S^y) \, \cos\phi \, \widetilde\rho_2 (b) \right] ,
\label{j_plus_rho}
\\[2ex]
\widetilde\rho_2 (b) &\equiv& \frac{\partial}{\partial b} 
\left[ \frac{\rho_2(b)}{2 M_N} \right] ,
\label{rho_2_tilde_def}
\end{eqnarray}
where $(...)$ hides a trivial factor reflecting the normalization 
of states (see Ref.~\cite{Granados:2013moa} for details);
$\cos\phi \equiv b^x/b$ is the cosine of the azimuthal angle, 
and $S^y = \pm 1/2$ the spin projection in the $y$--direction in the 
nucleon rest frame (see Fig.~\ref{fig:interpretation}).
The function $\rho_1(b)$ describes the spin--independent 
part of the current; the function $\cos\phi \, \widetilde\rho_2 (b)$
describes the spin--dependent part of the current 
in a transversely polarized nucleon.
%
% FIGURE
%

The spin--dependent part of the current
Eq.~(\ref{j_plus_rho}) changes sign between negative and positive
values of $b^x$, or ``left'' and ``right'' positions when looking 
at the nucleon along the negative $z$--direction (down from $z = +\infty$;
see Fig.~\ref{fig:interpretation}). The density $\widetilde\rho_2 (b)$
can thus be interpreted as the left--right asymmetry of the 
$J^+$ current in a nucleon polarized in the positive $y$ direction.
This interpretation has a natural connection with composite models
of nucleon structure, where polarization in the $y$--direction
generally induces a convection motion of the constituents around the
$y$--axis. As a result, the observer sees the charged constituents 
on the left side as ``blue--shifted'' (larger plus momentum), and
those on the right ride as ``red--shifted'' (smaller plus momentum),
compared to the unpolarized case \cite{Burkardt:2002hr}. 
We shall refer to this interpretation 
in our discussion of the peripheral chiral component in 
Sec.~\ref{subsec:transverse_polarization}. We note that Eq.~(\ref{j_plus_rho}) 
and its interpretation can be generalized to the case of arbitrary 
nucleon polarization states in the rest frame, including non--diagonal 
transitions; see Ref.~\cite{Granados:2013moa} for details.

The electromagnetic current matrix element and the transverse densities 
have two isospin components. In the following we are concerned with the
isovector component
\be
\langle N | J | N \rangle^V &\equiv& 
{\textstyle\frac{1}{2}} \left[ \langle p | J | p \rangle 
- \langle n | J | n \rangle \right] ,
\hspace{2em}
\rho_{1,2}^V \;\;\equiv \;\;
{\textstyle\frac{1}{2}} (\rho_{1, 2}^p - \rho_{1, 2}^n) .
\label{isospin}
\ee
The isoscalar component is defined by the same expression with the $+$ sign. 
\section{Chiral dynamics in current matrix element}
\label{sec:current_matrix_element}
\subsection{Peripheral chiral processes}
\label{subsec:peripheral_processes}
\begin{figure}[t]
\includegraphics[width=.62\textwidth]{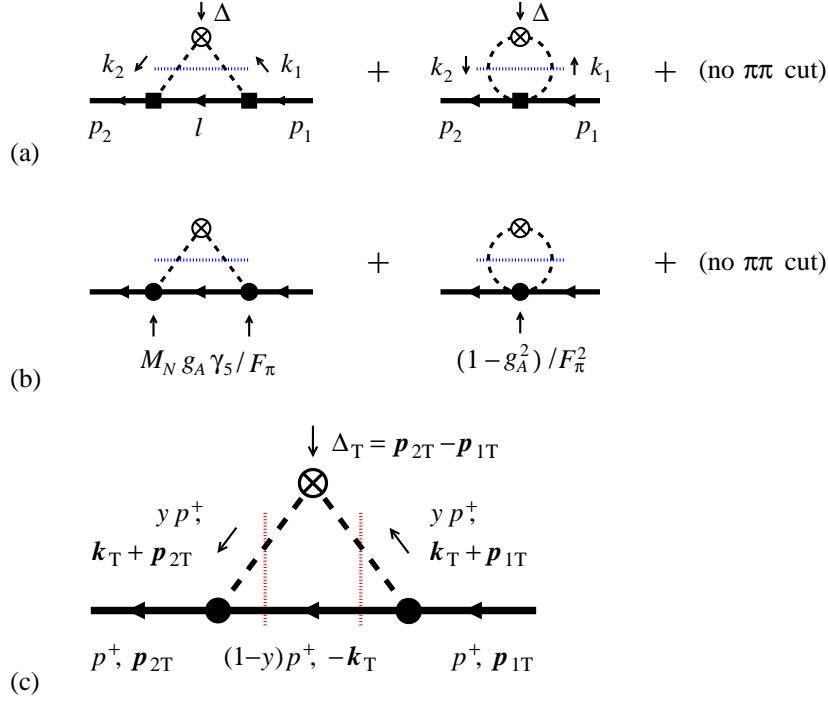}
\caption[]{(a) Feynman diagrams of LO chiral EFT processes contributing 
to the the two--pion cut of the isovector nucleon form factors and the 
peripheral transverse densities, Eq.~(\ref{me_chiral}). The squares denote 
the vertices of the relativistic chiral Lagrangian (axial vector $\pi NN$ 
coupling, $\pi\pi NN$ contact coupling). 
(b) Equivalent representation Eqs.~(\ref{me_equivalent})--(\ref{me_contact}).
The circles denote the pseudoscalar $\pi NN$ coupling and the effective
contact coupling $\propto (1 - g_A^2)$. (c) Light--front representation of 
the triangle diagram. The original nucleon makes a transition to a 
pion--nucleon intermediate state that couples to the current. The plus 
and transverse momenta are indicated.}
\label{fig:diag}
\end{figure}

At distances $b = O(M_\pi^{-1})$ the transverse densities 
are governed by universal chiral dynamics and can be calculated using 
methods of chiral EFT \cite{Strikman:2010pu,Granados:2013moa}. 
The isovector charge and magnetization densities in this region arise 
from chiral processes in which the current couples to the nucleon through 
exchange of a two--pion system in the $t$--channel. At LO these are
the processes described by the Feynman diagrams of Fig.~\ref{fig:diag}a,
where the vertices denote the pion--nucleon couplings of the LO relativistic
chiral Lagrangian \cite{Becher:1999he}. They produce densities of the form
\beq
\rho_{1, 2}^V(b) \;\; \sim \;\; P_{1, 2}(M_\pi, M_N; b) \; \exp(-2M_\pi b),
\label{large_b_general}
\eeq
where the the exponential decay is determined by the minimal mass of the 
exchanged system, $2 M_\pi$; the pre-exponential factors $P_{1, 2}$ is 
determined by the coupling of the exchanged system to the nucleon and 
exhibits a rich structure due to its dependence on the two scales, $M_\pi$ 
and $M_N$. Diagrams in which the current couples directly to the nucleon, 
or to a pion--nucleon vertex, produce contributions to the densities with 
range $O(M_N^{-1})$, or terms $\propto \delta^{(2)}(\bm b)$, and do not 
need to be considered in the calculation of the densities at 
$b = O(M_\pi^{-1})$.\footnote{The diagrams in which the current couples 
directly to the nucleon or to a pion--nucleon vertex renormalize the charge 
in the center of the nucleon, to compensate for the peripheral charge 
density produced by the two--pion exchange diagrams and ensure overall 
charge conservation.} In Ref.~\cite{Granados:2013moa} the densities at 
$b = O(M_\pi^{-1})$ resulting from the diagrams of Fig.~\ref{fig:diag}a
were computed in a dispersive representation, where the densities are
expressed as integrals of the imaginary parts of the form factors
on the cut at $t > 4 M_\pi^2$, and the integral extends over 
the parametric region $t - 4 M_\pi^2 = O(M_\pi^2)$.
%
% FIGURE
%

Now we want to represent the chiral dynamics generating the peripheral 
densities as actual processes evolving in light--front time $x^+$, and to
express the densities in terms of the light--front wave functions of the 
chiral $\pi N$ system. This could be done by solving the dynamical
problem of chiral EFT directly using light--front time--ordered perturbation 
theory \cite{Brodsky:1997de}. A more convenient approach is to take the known 
chiral EFT result in the relativistically invariant formulation and 
rewrite it such that it corresponds to the overlap of light--front wave 
functions. This approach maintains the connection with the invariant
formulation and naturally generates also the instantaneous terms 
(zero modes) that require special considerations in the 
time--ordered approach. In the LO approximation the invariant chiral EFT 
result for the isovector nucleon current matrix element 
Eq.~(\ref{me_general}) at the position $x = 0$ 
is \cite{Gasser:1987rb,Bernard:1992qa,Kubis:2000zd,Kaiser:2003qp}
(the specific form here was derived in Ref.~\cite{Granados:2013moa})
\be
\langle N_2 | \, J^\mu (0) \, | N_1 \rangle^V
&=& 
\int\!\frac{d^4 k}{(2\pi)^4} \; 
\frac{i \, k^\mu}{(k_1^2 - M_\pi^2 + i0) (k_2^2 - M_\pi^2 + i0)} 
\left[ \frac{g_A^2}{F_\pi^2} 
\; \frac{\bar u_2 \hat k_2 \gamma_5 (\hat l + M_N) \hat k_1 \gamma_5 u_1}
{l^2 - M_N^2 + i0} 
+ \frac{1}{F_\pi^2} \;
\bar u_2 \hat k u_1
\right]
\nonumber \\[2ex]
&+& \textrm{(diagrams without $\pi\pi$ cut)} ,
\label{me_chiral}
\ee
where 
\beq
k_{1, 2} \;\; = \;\; k \mp \Delta/2 
\label{k_12_def}
\eeq
are the 4--momenta of the pions coupling to the vector current, and the 
average 4--momentum $k$ was chosen as integration variable. 
The first term in the brackets results 
from the triangle diagram in Fig.~\ref{fig:diag}a 
and is proportional to the squared $\pi NN$ coupling $g_A^2/F_\pi^2$;
it involves the intermediate nucleon propagator with 4--momentum
\be
l &\equiv& p_1 - k_1 \;\; = \;\; p_2 - k_2 
\label{l_def}
\ee
(we use the notation $\hat k_{1,2} \equiv k_{1,2}^\mu \gamma_\mu, \,
\hat l \equiv l^\mu \gamma_\mu$).
The second term results from the contact diagram in Fig.~\ref{fig:diag}a and 
is proportional to the $\pi\pi NN$ contact coupling in the 
chiral Lagrangian, $1/F_\pi^2$. As explained above, Eq.~(\ref{me_chiral}) 
shows only the contribution from the $\pi\pi$ cut diagrams that
contribute to te peripheral density. 

The first term in Eq.~(\ref{me_chiral}) contains a piece in which the 
pole of the nucleon propagator cancels, and which is of the same form 
as the second term. For deriving a wave function overlap representation 
it is important to extract this piece and combine it with the second term.
Expressing the pion momenta as $k_{1, 2} = p_{1, 2} - l$, 
cf.~Eq.~(\ref{l_def}), using the anticommutation relations between the 
gamma matrices, and the Dirac equation for the external nucleon spinors, 
$\hat p_1 u_1 = M_N u_1$ and $\bar u_2 \hat p_2 = \bar u_2 M_N$, we rewrite 
the bilinear form in the numerator of the first term in 
Eq.~(\ref{me_chiral}) as 
\be
\bar u_2 \hat k_2 \gamma_5 (\hat l + M_N) \hat k_1 \gamma_5 u_1 
&=& - 4 M_N^2 \bar u_2 \gamma_5 (\hat l + M_N) \gamma_5 u_1 
\; - \; (l^2 - M_N^2) [ \bar u_2 \hat k u_1 \; 
+ \; (\textrm{terms even in $k$})].
\label{bilinear_simplified}
\ee
In the second term the factor $l^2 - M_N^2$ cancels the pole of the 
nucleon propagator. Moreover, the terms even under $k \rightarrow -k$ 
in the parentheses integrate to zero because after canceling the nucleon
pole the remaining integrand in Eq.~(\ref{me_chiral}) is odd under 
$k \rightarrow -k$. Altogether, we can thus replace the
terms in the bracket in Eq.~(\ref{me_chiral}) by
\be
\left[\phantom{\frac{0^0}{0^0}} \hspace{-1em} \ldots \; \right] \;\; = \;\;
\left[ -\frac{4 M_N^2 g_A^2}{F_\pi^2} 
\; \frac{\bar u_2 \gamma_5 (\hat l + M_N) \gamma_5 u_1}{l^2 - M_N^2 + i0} 
\; + \; \frac{1 - g_A^2}{F_\pi^2} \; \bar u_2 \hat k u_1
\right] .
\ee
The peripheral chiral EFT contribution to the current matrix element
can therefore be represented as the sum of two terms (see
Fig.~\ref{fig:diag}b)
\be
\langle N_2 | \, J^\mu (0) \, | N_1 \rangle^V
&=& \langle \ldots \rangle^V_{\rm interm}
\;\; + \;\; 
\langle \ldots \rangle^V_{\rm contact} ,
\label{me_equivalent}
\\[2ex]
\langle N_2 | \, J^\mu (0) \, | N_1 \rangle^V_{\rm interm}
&\equiv& \frac{4 M_N^2 g_A^2}{F_\pi^2} \;
\int\!\frac{d^4 k}{(2\pi)^4} \; 
\; \frac{(-i) \, k^\mu \, \bar u_2 \gamma_5 (\hat l + M_N) \gamma_5 u_1}
{(k_1^2 - M_\pi^2 + i0) (k_2^2 - M_\pi^2 + i0) (l^2 - M_N^2 + i0)} ,
\label{me_onshell}
\\[1ex]
\langle N_2 | \, J^\mu (0) \, | N_1 \rangle^V_{\rm contact}
&\equiv& \frac{1 - g_A^2}{F_\pi^2} \;
\int \! \frac{d^4 k}{(2\pi )^4} 
\; \frac{i k^\mu \, \bar u_2 \hat k u_1}{(k_1^2 - M_\pi^2 + i0)
(k_2^2 - M_\pi^2 + i0)} .
\label{me_contact}
\ee
The first term, Eq.~(\ref{me_onshell}), contains the intermediate nucleon 
propagator and is identical in form to the pion loop graph with the 
usual \textit{pseudoscalar} $\pi NN$ vertex with effective coupling 
$g_{\pi NN} = M_N g_A/F_\pi$. This will allow us to derive a wave function 
overlap representation for this term with the pseudoscalar vertex, 
which is free of the ambiguities of the momentum--dependent
axial vector coupling. The second term, Eq.~(\ref{me_contact}), represents 
an effective contact term, combining the explicit $\pi\pi NN$ 4--point vertex
in the chiral Lagrangian with the ``non-propagating'' part of the triangle
diagram. The appearance of the combination $1 - g_A^2$ indicates that this 
term expresses internal structure of the nucleon (for a pointlike Dirac 
fermion $g_A = 1$) and that its contribution is numerically small;
cf.\ the discussion in Sec.~\ref{subsec:contact}. The two terms 
in the current matrix element thus have distinct physical meaning and 
will be discussed separately in the following. We emphasize that the 
decomposition Eqs.~(\ref{me_equivalent})--(\ref{me_contact})
is obtained by identical rewriting of the original Feynman integrals 
and does not involve additional approximations. 
\subsection{Overlap representation}
\label{subsec:overlap}
The intermediate--nucleon term of the current matrix element can be 
represented as an overlap integral of light--front wave functions. 
We derive this representation through a suitable three--dimensional
reduction of the Feynman integral Eq.~(\ref{me_onshell}). 
To this end we go to a class of frames where the 
momentum transfer has only transverse components, 
cf.\ Eq.~(\ref{Delta_transverse}), such that (see Fig.~\ref{fig:diag}c)
\be
p_1^+ \; = \; p_2^+ \; \equiv \; p^+, 
\hspace{2em} \bm{p}_{2T} - \bm{p}_{1T} \; = \; \bm{\Delta}_T,
\hspace{2em} p_{1}^- \; = \; \frac{M_N^2 + \bm{p}_{1T}^2}{p^+} ,
\hspace{2em} p_{2}^- \; = \; \frac{M_N^2 + \bm{p}_{2T}^2}{p^+} .
\label{transverse_frame}
\ee
The plus component $p^+ > 0$ is a free parameter, whose choice selects a
particular frame in a class of frames related by longitudinal boosts.
Likewise, the overall transverse momentum remains unspecified; only the
difference $\bm{p}_{2T} - \bm{p}_{1T}$ is required to be equal to the
momentum transfer $\bm{\Delta}_T$.
We introduce light--front components of the loop momentum 
$k^\pm \equiv k^0 \pm k^z$ and $\bm{k}_T \equiv (k^x, k^y)$,
\be
\int d^4 k &=& \frac{1}{2} \int dk^+ \int dk^- \int d^2 k_T ,
\ee
and express $k^+$ in terms of the boost--invariant pion momentum
fraction $y$,
\be
k^+ \; = \; y p^+ .
\ee
The integrand of Eq.~(\ref{me_onshell}) has simple poles in $k^-$, at 
the values determined by the mass shell conditions for the pion and 
nucleon 4--momenta. The two poles of the pion propagators and the
one of the nucleon propagator lie on opposite sides of the real axis
if $0 < y < 1$. The integral over $k^-$ can thus be taken by closing 
the contour around the nucleon pole. At the nucleon pole the pion
virtualities take the values
\be
k_{1}^2 - M_\pi^2 &=&  
-\frac{(\bm{k}_T + \bar y\bm{p}_{1T})^2}{\bar y} 
- \frac{y^2 M_N^2}{\bar y} - M_\pi^2 \;\; < \;\; 0,
\\[1ex]
k_{2}^2 - M_\pi^2 &=&  
-\frac{(\bm{k}_T + \bar y\bm{p}_{2T})^2}{\bar y} 
- \frac{y^2 M_N^2}{\bar y} - M_\pi^2 \;\; < \;\; 0,
\ee
where
\beq
\bar y \;\; \equiv \;\; 1 - y.
\eeq
These virtualities can be related to the invariant mass differences 
between states in the light--front time--ordered formulation, 
in which the external nucleon makes a transition to an intermediate 
$\pi N$ state and back (see Fig.~\ref{fig:diag}c). 
The invariant mass difference for the 
transition from the initial nucleon state with plus momentum $p^+$ and 
transverse momentum $\bm{p}_{1T}$ to a pion with $y p^+$ and 
$\bm{k}_T + \bm{p}_{1T}$ and a nucleon with $\bar y p^+$ and $-\bm{k}_T$ 
is given by
\be
\Delta\mathcal{M}^2 (y, \bm{k}_T, \bm{p}_{1T}) &\equiv&
\frac{(\bm{k}_T + \bm{p}_{1T})^2 + M_\pi^2}{y} + 
\frac{\bm{k}_T^2 + M_N^2}{\bar y} - M_N^2 - \bm{p}_{1T}^2
\label{invariant_mass_explicit_orig}
\\[2ex]
&=&
\frac{(\bm{k}_T + \bar y \bm{p}_{1T})^2 + M_\pi^2}{y} + 
\frac{(\bm{k}_T + \bar y\bm{p}_{1T})^2 + M_N^2}{\bar y} - M_N^2 .
\label{invariant_mass_explicit}
\ee
The invariant mass difference for the transition from the final nucleon 
state with $p^+$ and $\bm{p}_{2T}$ to a pion with $y p^+$ and 
$\bm{k}_T + \bm{p}_{2T}$ and a nucleon with
$\bar y p^+$ and $-\bm{k}_T$ is given by the same expressions
with $\bm{p}_{1T} \rightarrow \bm{p}_{2T}$.
It is easy to see that
\be
- \frac{k_{1}^2 - M_\pi^2}{y} &=&  
\Delta\mathcal{M}^2 (y, \bm{k}_T, \bm{p}_{1T}) ,
\label{virtuality_invariant_mass_1}
\\
- \frac{k_{2}^2 - M_\pi^2}{y} &=&  
\Delta\mathcal{M}^2 (y, \bm{k}_T, \bm{p}_{2T}) .
\label{virtuality_invariant_mass_2}
\ee
Equations~(\ref{virtuality_invariant_mass_1}) and 
(\ref{virtuality_invariant_mass_2}) allow us to interpret the pion 
propagators in the Feynman integral as invariant mass denominators.
[The origin of the expression Eq.~(\ref{invariant_mass_explicit_orig})
and its connection with the ``energy denominator'' in light--front 
time--ordered perturbation theory are explained in 
Appendix~\ref{app:time-ordered}; this information is not needed
for the calculations performed here but important for general
understanding.]

Further, at the pole of the nucleon propagator the numerator of 
Eq.~(\ref{me_onshell}) can be factorized. Since at the pole the 
4--momentum $l$ is on the mass shell, the matrix $\hat l + M_N$ 
coincides with the projector on physical nucleon spin states
and can be represented as
\be
(M_N + \hat l)_{\rm on-shell} 
&=& \sum_{\sigma = \pm 1/2} u(l, \sigma) \, \bar u(l, \sigma) ,
\label{spin_projector}
\ee
where $u(l, \sigma)$ is a set of nucleon 4--spinors; the choice of
polarization states will be specified below. We thus can write the
bilinear form in the numerator as [reverting to the full notation 
$u_1 \equiv u(p_1, \sigma_1)$ and $u_2 \equiv u(p_2, \sigma_2)$]
\be
- \bar u_2 \gamma_5 (\hat l + M_N) \gamma_5 u_1|_{\rm on-shell} &=&
\sum_{\sigma = \pm 1/2} u(p_2, \sigma_2) i \gamma_5 \bar u (l, \sigma)
\; \bar u (l, \sigma) i \gamma_5 u(p_1, \sigma_1) .
\label{numerator_factorized}
\ee
Here it is understood that the on-shell 4--momentum $l$ is expressed in 
terms of the remaining integration variables $y$ and $\bm{k}_T$,
\beq
l^+ = \bar y p^+, \hspace{2em} l^- \; = \; 
\frac{|\bm{k}_T|^2 + M_N^2}{\bar y p^+},
\hspace{2em} \bm{l}_T \; = \; - \bm{k}_T .
\eeq
The bilinear forms appearing on the right--hand side of 
Eq.~(\ref{numerator_factorized}) can be related to the vertex functions 
for an $N \rightarrow \pi N$ transition with specified on-shell 
nucleon momenta and spin and its complex conjugate. 
Defining the pseudoscalar vertex function for the transition 
from the initial nucleon 
state with momentum $p^+$ and $\bm{p}_{1T}$ and spin $\sigma_1$ to a nucleon 
with momentum $\bar y p^+$ and $-\bm{k}_T$ and spin $\sigma$ as
\beq
\Gamma (y, \bm{k}_T, \bm{p}_{1T}; \sigma, \sigma_1) \;\; \equiv \;\;
\frac{g_A M_N}{F_\pi} \; \bar u(l, \sigma) i \gamma_5 u (p_1, \sigma_1) ,
\label{Gamma}
\eeq
the vertex for the transition to the final state is given by
\beq
\frac{g_A M_N}{F_\pi} \;
u(p_2, \sigma_2) i \gamma_5 \bar u (l, \sigma) \;\; = \;\;
\frac{g_A M_N}{F_\pi} \;
[\bar u (l, \sigma) i \gamma_5 u(p_2, \sigma_2)]^\ast = 
\Gamma^\ast (y, \bm{k}_T, \bm{p}_{2T}; \sigma, \sigma_2 ) ,
\label{Gamma_conjugate}
\eeq
and multiplying Eq.~(\ref{numerator_factorized}) by the squared
coupling constant we obtain
\be
\frac{g_A^2 M_N^2}{F_\pi^2} \;
\bar u_2 \gamma_5 (\hat l + M_N) \gamma_5 u_1|_{\rm on-shell} &=&
\sum_{\sigma = \pm 1/2} 
\Gamma^\ast (y, \bm{k}_T, \bm{p}_{2T}; \sigma, \sigma_2 )
\Gamma (y, \bm{k}_T, \bm{p}_{1T}; \sigma, \sigma_1) .
\ee
We now \textit{define} the light--front wave function of the 
initial state in the process of Fig.~\ref{fig:diag}c as
\be
\Psi (y, \bm{k}_T, \bm{p}_{1T}; \sigma, \sigma_{1}) 
&\equiv& 
\frac{\Gamma (y, \bm{k}_T, \bm{p}_{1T}; \sigma, \sigma_{1})}
{\Delta\mathcal{M}^2(y, \bm{k}_T, \bm{p}_{1T})} ;
\label{psi_def}
\ee
the wave function for the final state is given by the same
expression with $\bm{p}_{1T} \rightarrow \bm{p}_{2T}$ and
$\sigma_1 \rightarrow \sigma_2$ (i.e., it is the same function
but evaluated at a different argument). It is then
straightforward to compute the integral over $k^-$,
and the intermediate--nucleon part of the current 
matrix element Eq.~(\ref{me_onshell}) becomes
\be
\langle N_2 | \, J^+ (0) \, | N_1 \rangle^V_{\rm interm}
&\equiv&  
\langle N (p^+, \bm{p}_{2T}, \sigma_2 ) 
| \, J^+ (0) \, | N (p^+, \bm{p}_{1T}, \sigma_1 )  
\rangle^V_{\rm interm}
\nonumber
\\[1ex]
&=& \frac{(2 p^+)}{2\pi} \int\frac{dy}{y\bar y} \int \frac{d^2 k_T}{(2\pi)^2}
\sum_{\sigma} \Psi^\ast (y, \bm{k}_T, \bm{p}_{2T}; \sigma, \sigma_2) \;
\Psi (y, \bm{k}_T, \bm{p}_{1T}; \sigma, \sigma_1) .
\label{me_onshell_overlap}
\ee
The original Feynman integral is represented as an overlap integral of the 
light--front wave functions describing the transition from the initial 
nucleon state $N_1$ to a $\pi N$ intermediate state and back to the final 
nucleon state $N_2$. Equation~(\ref{me_onshell_overlap}) will be our
starting point for the analysis of the transverse densities.

The wave function Eq.~(\ref{psi_def}) is defined in terms of the vertex
function obtained from the chiral Lagrangian and the invariant mass
difference of the $N \rightarrow \pi N$ transition. 
Appendix~\ref{app:time-ordered} shows that this object is identical
to the traditional light--front wave function, 
defined as the transition matrix element 
between the initial nucleon state and the intermediate pion--nucleon
state in light--front time--ordered perturbation theory.
Regarding isospin the wave function Eq.~(\ref{psi_def}) 
is normalized such that it describes 
the transition $p \rightarrow \pi^0 + p$, for which the coupling 
is $g_{\pi^0 pp} \equiv g_{\pi NN} = g_A M_N/ F_\pi$. For the transitions
$p \rightarrow \pi^+ + n$ and $n \rightarrow \pi^- + p$, which actually 
contribute to the isovector electromagnetic current matrix element, 
the couplings follow from isospin invariance and are 
$g_{\pi^+ np} = g_{\pi^- pn} = 
\sqrt{2} \, g_{\pi NN}$. The isospin factor $\sqrt{2} \times \sqrt{2} = 2$ 
is included in the prefactor of Eq.~(\ref{me_onshell_overlap}).

As explained above, we are interested in the current matrix element only
in the (unphysical) region of momentum transfers in the vicinity of the 
two--pion threshold in the $t$--channel, $t - 4 M_\pi^2 = O(M_\pi^2)$, 
and consider the wave function representation 
Eq.~(\ref{me_onshell_overlap}) only in this parametric domain.
The restriction to this domain will appear naturally when going over 
to the coordinate representation and 
considering the region of transverse distances $b = O(M_\pi^{-1})$.
\section{Chiral light--front wave function}
\label{sec:wave_function}
\subsection{Nucleon spin states}
\label{subsec:spin}
To proceed with the evaluation of the overlap formula 
Eq.~(\ref{me_onshell_overlap}) we
need to specify the nucleon spin states and obtain explicit
expressions for the vertex functions Eq.~(\ref{Gamma}) and
(\ref{Gamma_conjugate}). Equation~(\ref{me_onshell_overlap})
can be evaluated with any choice of nucleon spin states
(external and internal); the resulting expressions and
their interpretation depend on the choice, of course. It is
natural to choose the nucleon spin states as light--front helicity
states \cite{Brodsky:1997de}. For a nucleon state with light--front
momentum $p^+$ and $\bm{p}_T$ the
light--front helicity spinors are obtained by subjecting the Dirac
spinors in the rest frame [$p^+(\textrm{RF}) = M_N, \,
\bm{p}_T(\textrm{RF}) = 0$] first to a longitudinal boost from $M_N$
to $p^+$, and then to a transverse boost from $0$ to $\bm{p}_T$.
The spinors thus defined are invariant under longitudinal boosts and
transform in a simple manner under transverse boosts. An explicit
representation of the light--front helicity spinors is
\cite{Brodsky:1997de,Leutwyler:1977vy}
\beq
u(p, \sigma) \;\; \equiv \;\; u(p^+, \bm{p}_{T}, \sigma ) \;\; = \;\; 
\frac{1}{\sqrt{2 p^+}} \left[ p^+ \gamma^- 
+ (M_N - \bm{\gamma}_T \cdot \bm{p}_{T} ) \gamma^+ \right] 
\left( \begin{array}{c} \chi (\sigma ) \\[1ex] 0 \end{array} \right) ,
\label{spinor_boosted}
\eeq
where $p$ denotes the on-shell 4--momentum vector ($p^2 = M_N^2$),
$\sigma = \pm 1/2$, and $\chi (\sigma)$ are rest frame 2--spinors 
for polarization in the positive and negative $z$--direction,
\beq
\chi(\sigma = 1/2) \; = \; 
\left( \begin{array}{c} 1 \\ 0 \end{array}\right),
\hspace{2em}
\chi(\sigma = -1/2) \; = \; 
\left( \begin{array}{c} 0 \\ 1 \end{array}\right) .
\label{chi_along_z}
\eeq
The spinors are normalized such that $\bar u u = 2 M_N$ and satisfy
the completeness relation Eq.~(\ref{spin_projector}). Using these
spinors to evaluate the pseudoscalar vertex Eq.~(\ref{Gamma})
one gets\footnote{The sign of the three--dimensional expressions 
for the vertex function depends on the convention for the matrix 
$\gamma^5$. We use the Bjorken--Drell convention 
$\gamma^5 = i \gamma^0 \gamma^1 \gamma^2 \gamma^3$.}
\be
\Gamma (y, \bm{k}_T, \bm{p}_{1T}; \sigma, \sigma_{1})
&=& \frac{2i g_A M_N}{F_\pi \sqrt{\bar{y}}}
\left[ y M_N \, S^z (\sigma, \sigma_{1}) + 
\left(\bm{k}_T + \bar y \bm{p}_{1T} \right) 
\cdot \bm{S}_T(\sigma, \sigma_{1}) 
\right] ,
\label{vertex_explicit}
\ee
and similarly for the vertex with $\bm{p}_{2T}$ and $\sigma_2$.
Here $S^z$ and $\bm{S}_T \equiv (S^x, S^y)$ are the components of the
3--vector characterizing the spin transition matrix element in the
rest frame
\be
S^i (\sigma, \sigma_1) &\equiv& \chi^\dagger (\sigma)
({\textstyle\frac{1}{2}} \sigma^i ) \chi (\sigma_1) 
\hspace{2em} (i = x, y, z),
\label{S_vector}
\ee
where $\sigma^i$ are the Pauli matrices.
Use of this compound variable results in a compact representation of the 
light--front spin structure in close correspondence to non-relativistic 
quantum mechanics.

\begin{figure}[t]
\includegraphics[width=.68\textwidth]{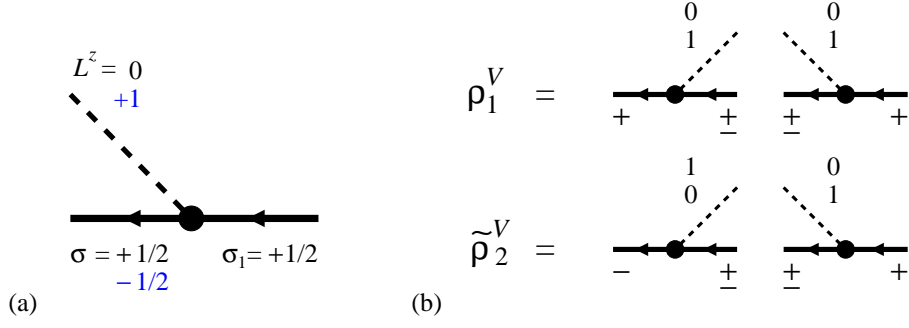}
\caption[]{(a) Spin structure of the $N \rightarrow \pi N$ light--front
wave function in chiral EFT.
The light--front helicity corresponds to the $z$--projection of the
nucleon spin in the rest frame. The helicity--conserving component
has orbital angular momentum projection $L^z = 0$, the helicity--flip
component has $L^z = 1$. (b) Wave function overlap representation of 
the transverse densities $\rho_1^V(b)$ and $\widetilde\rho_2^V(b)$
in the light--front helicity representation,
Eq.~(\ref{rho_overlap}).}
\label{fig:diag_wf}
\end{figure}

The vertex Eq.~(\ref{vertex_explicit}) contains two structures with 
different orbital angular momentum (see Fig.~\ref{fig:diag_wf}). 
The first term on the right--hand side is diagonal
in the light--front helicity, because 
\beq
S^z (\sigma, \sigma_{1}) 
\;\; = \;\; \sigma \; \delta(\sigma, \sigma_{1})
\eeq
when the rest frame spinors are eigenspinors of $\sigma^z$, cf.\
Eq.~(\ref{chi_along_z}). It describes a transition $N \rightarrow \pi N$ 
in which the nucleon light--front 
helicity is preserved and the $\pi N$ state has orbital 
angular momentum projection $L^z = 0$. The second term is off-diagonal 
in light--front
helicity, because $\sigma^x$ and $\sigma^y$ have only off-diagonal elements.
It corresponds to a transition in which the nucleon helicity is flipped and 
the $\pi N$ system has orbital angular momentum projection $L^z = 1$. This 
is immediately obvious from the fact that this term is proportional to the
transverse momentum $\bm{k}_T + \bar y \bm{p}_{1T}$, which transforms 
as 2--dimensional vector under rotations around the $z$--axis.
%
% FIGURE
%
\subsection{Transverse rest frame} 
Equation~(\ref{me_onshell_overlap}) represents the current matrix element as
an overlap integral of the light--front wave functions of the initial and 
final nucleon states with overall transverse momenta $\bm{p}_{1T}$
and $\bm{p}_{2T}$. For further analysis is will be convenient
to express these wave functions in terms of the wave function at zero overall
transverse momentum (transverse rest frame), such that the overlap integral
becomes a quadratic form in a single function. This can be accomplished using
the transformation properties under transverse boosts. As can be seen
from Eqs.~(\ref{psi_def}), (\ref{invariant_mass_explicit}) and 
(\ref{vertex_explicit}), the wave function at overall transverse momentum 
$\bm{p}_{1T}$ is related to that at zero transverse momentum by
\be
\Psi (y, \bm{k}_T, \bm{p}_{1T}; \sigma, \sigma_1) &=& 
\Psi (y, \bm{k}_T + \bar y \bm{p}_{1T}, \, \bm{0}; \sigma, \sigma_1) 
\;\; \equiv \;\;
\Psi (y, \bm{k}_T + \bar y \bm{p}_{1T}; \sigma, \sigma_1) .
\ee
The expression for the rest frame wave function can be obtained 
by setting $\bm{p}_{1T} = 0$ in Eqs.~(\ref{invariant_mass_explicit})
and (\ref{vertex_explicit}). For reference we quote the explicit formulas:
\be
\Psi (y, \widetilde{\bm{k}}_T; \sigma, \sigma_1)
&\equiv& 
\frac{\Gamma (y, \widetilde{\bm{k}}_T; \sigma, \sigma_1)}
{\Delta\mathcal{M}^2(y, \widetilde{\bm{k}}_T)} ,
\label{psi_restframe}
\\[1ex]
\Delta\mathcal{M}^2 (y, \widetilde{\bm{k}}_T) &\equiv&
\frac{\widetilde{\bm{k}}_T^2 + M_\pi^2}{y} + 
\frac{\widetilde{\bm{k}}_T^2 + M_N^2}{\bar y} - M_N^2 ,
\label{invariant_mass_restframe}
\\[1ex]
\Gamma (y, \widetilde{\bm{k}}_T; \sigma, \sigma_1)
&=& \frac{2i g_A M_N}{F_\pi \sqrt{\bar{y}}}
\left[ y M_N \, S^z (\sigma, \sigma_1) + 
\widetilde{\bm{k}}_T \cdot \bm{S}_T(\sigma, \sigma_1) 
\right] ,
\label{vertex_restframe}
\ee
where we use $\widetilde{\bm{k}}_T$ to denote the transverse 
momentum argument. 
Similar formulas apply to the outgoing wave function with transverse 
momentum $\bm{p}_{2T}$ and spin $\sigma_2$.
The current matrix element Eq.~(\ref{me_onshell_overlap}) 
can therefore equivalently be expressed in terms of the rest frame 
wave function,
\be
\langle N_2 | \, J^+ (0) \, | N_1 \rangle^V_{\rm interm}
&=& 
\frac{(2 p^+)}{2\pi} \int\frac{dy}{y\bar y} \int \frac{d^2 k_T}{(2\pi)^2}
\; \sum_{\sigma} 
\Psi^\ast (y, \bm{k}_T + \bar y \bm{p}_{2T}; \sigma , \sigma_2) \;
\Psi (y, \bm{k}_T + \bar y \bm{p}_{1T}; \sigma, \sigma_1) .
\label{me_onshell_overlap_restframe}
\ee
We shall use this expression in our theoretical studies in the following.
\subsection{Coordinate representation}
It is instructive to study the rest frame light--front wave function in the
transverse coordinate representation. The coordinate--space wave function
allows us to identify the parametric regime of peripheral distances where
chiral dynamics is valid and to calculate the transverse densities
directly in coordinate space. We define the coordinate--space wave
function as the transverse Fourier transform of the momentum--space wave 
function at fixed plus momentum fraction $y$,
\be
\Phi(y, \bm{r}_T, \sigma, \sigma_1) &\equiv&
\int \frac{d^2\widetilde{k}_{T}}{(2\pi)^2} \; e^{i \bm{r}_T \cdot
\bm{\widetilde{k}}_T} \; \Psi (y, \bm{\widetilde{k}}_T; \sigma, \sigma_1) .
\label{psi_coordinate}
\ee
The vector $\bm{r}_T$ is the difference in the transverse positions of the 
$\pi$ and $N$ (relative transverse coordinate), such that the wave function 
describes the physical transverse size distribution of the $\pi N$ system.
Because of the spin structure of the vertex Eq.~(\ref{psi_restframe})
the coordinate--space wave function can be expressed in terms of two 
transverse radial wave functions (i.e., scalars with respect 
to rotations around the $z$--axis),
\be
\Phi(y,\bm{r}_T, \sigma, \sigma_1) \;\; = \;\; 
-2i S^z(\sigma, \sigma_1) \; U_0(y, r_T) \; + \; \frac{2 \, \bm{r}_T
\cdot \bm{S}_T (\sigma, \sigma_1)}{r_T} \; U_1(y, r_T) ,
\label{psi_coordinate_decomposition}
\ee
where $r_T \equiv |\bm{r}_T|$ is the modulus of the
transverse coordinate. Following Sec.~\ref{subsec:spin} 
these are the components with orbital angular momentum projection
$L^z = 0$ and 1. The Fourier integral is easily calculated
by writing the invariant mass in the denominator of the momentum--space
wave function, Eq.~(\ref{invariant_mass_restframe}), in the form
\be
\Delta\mathcal{M}^2 &=& \frac{\bm{\widetilde{k}}_T^2 + M_T^2}{y \bar y} ,
\label{invariant_mass_transverse}
\\[2ex]
M_T &\equiv& M_T(y) \;\, \equiv \;\, 
\sqrt{\bar{y} M_\pi^2 +y^2 M_N^2} ,
\label{M_T_def}
\ee
which is the $y$--dependent effective mass governing the transverse 
momentum dependence. We obtain
\be
\left.
\begin{array}{r}
U_0(y,r_T) 
\\[2ex]
U_1(y,r_T)
\end{array}
\right\}
&=& \frac{g_A M_N \, y \sqrt{\bar{y}}}{2\pi F_\pi}
\left\{
\begin{array}{r}
y M_N \; K_0(M_T r_T)
\\[2ex]
M_T \; K_1(M_T r_T)
\end{array}
\right\} ,
\label{psi_0_1_def}
\ee
where $K_0$ and $K_1$ are the modified Bessel functions.
At large values of the argument they behave as
\beq
K_{0, 1} (M_T r_T) \;\; \sim \;\; \sqrt{\frac{\pi}{2}}
\; \frac{e^{-M_T r_T}}{\sqrt{M_T r_T}}
\hspace{2em} (M_T r_T \; \gg \; 1) .
\label{K01_asymptotic}
\eeq
The coordinate--space wave functions fall off exponentially at large 
transverse distances $r_T$, with a width that is given by the
transverse mass Eq.~(\ref{M_T_def}) and depends on the pion momentum 
fraction $y$. This behavior can directly be traced to the singularity 
of the momentum--space wave function at zero invariant mass, 
$\Delta \mathcal{M}^2 = 0$, which occurs at complex values of
the transverse momentum, $\bm{\widetilde{k}}_T^2 = - M_T^2$, 
cf.~Eq.~(\ref{invariant_mass_transverse}).

The parametric domain in which we are interested in the coordinate--space 
wave function is 
\beq
y \;\; = \;\; O(M_\pi / M_N),
\hspace{2em} 
r_T \;\; = \;\; O(M_\pi^{-1}) .
\label{parametric_chiral}
\eeq
In momentum space this corresponds to the region where the pion's 
light--front momentum components in the nucleon rest frame are
\beq
\widetilde{k}^+ \;\; = \;\; y M_N \;\; = \;\; O(M_\pi), 
\hspace{2em}
|\bm{\widetilde{k}}_T| \;\; = \;\; O(M_\pi),
\eeq
and also $\widetilde k^- = (|\bm{\widetilde{k}}_T|^2 
+ M_\pi^2)/\widetilde{k}^+ = O(M_\pi)$, such that
all components of the pion's 4--momentum are $O(M_\pi)$ (``soft pion'').
In this region chiral dynamics is applicable, and the approximations made
in evaluating the peripheral contributions to the current matrix element
are self-consistent. Equation~(\ref{M_T_def}) shows that
for momentum fractions $y = O(M_\pi / M_N)$
\beq
M_T(y) \;\; = \;\; O(M_\pi) \hspace{2em} [y = O(M_\pi / M_N)],
\eeq
so that the exponential range of the coordinate--space wave function is
indeed of the order $O(M_\pi^{-1})$, cf.~Eq.~(\ref{K01_asymptotic}).

We note that for $y = O(1)$ the effective mass Eq.~(\ref{M_T_def})
is $M_T = O(M_N)$, so that the range of the wave function
Eq.~(\ref{psi_0_1_def}) is $O(M_N^{-1})$. While the wave 
is still formally defined by Eq.~(\ref{psi_0_1_def}), it does not 
correspond to a chiral long--distance contribution in this case. 
This region does not contribute to the peripheral transverse densities, 
as the wave functions for $y = O(1)$ are exponentially small if the 
distance is kept at values $r_T = O(M_\pi^{-1})$. In the calculations
in Sec.~\ref{sec:densities} we can thus formally integrate up to $y = 1$
without violating the parametric restriction Eq.~(\ref{parametric_chiral}).

It is interesting to compare the parametric order of the light--front 
helicity--nonflip ($L^z = 0$) and flip ($L^z = 1$) components of the 
coordinate--space wave function in $M_\pi/M_N$.
Inspection of Eq.~(\ref{psi_0_1_def}) shows that for
$y = O(M_\pi/M_N)$ and $r_T = O(M_\pi)$, Eq.~(\ref{parametric_chiral}),
\beq
U_0/U_1 \;\; = \;\; O(1) .
\label{psi_0_psi_1_ratio}
\eeq
The helicity--nonflip and flip components are thus of the same order 
in the region of interest.\footnote{At exceptionally small pion momentum 
fractions $y \ll M_\pi /M_N$ that the helicity--nonflip component of
the wave function vanishes faster than the helicity--flip one,
$U_0 / U_1 \rightarrow 0$. This scenario is realized in the 
``molecular'' region described in Ref.~\cite{Granados:2013moa}.}
Regarding the numerical values we note that
\beq
U_0(y, r_T) \;\; < \;\; U_1 (y, r_T) 
\hspace{2em} (0 < y < 1, \; r_T > 0) ,
\label{psi_0_psi_1_inequality}
\eeq
because $M_T (y) > y M_N$, cf.~Eq.~(\ref{M_T_def}), and
$K_1 (z) > K_0(z)$ for all $z > 0$. The radial wave functions thus obey a
numerical inequality at all values of the argument.

\begin{figure}
\begin{tabular}{ll}
\includegraphics[width=.48\textwidth]{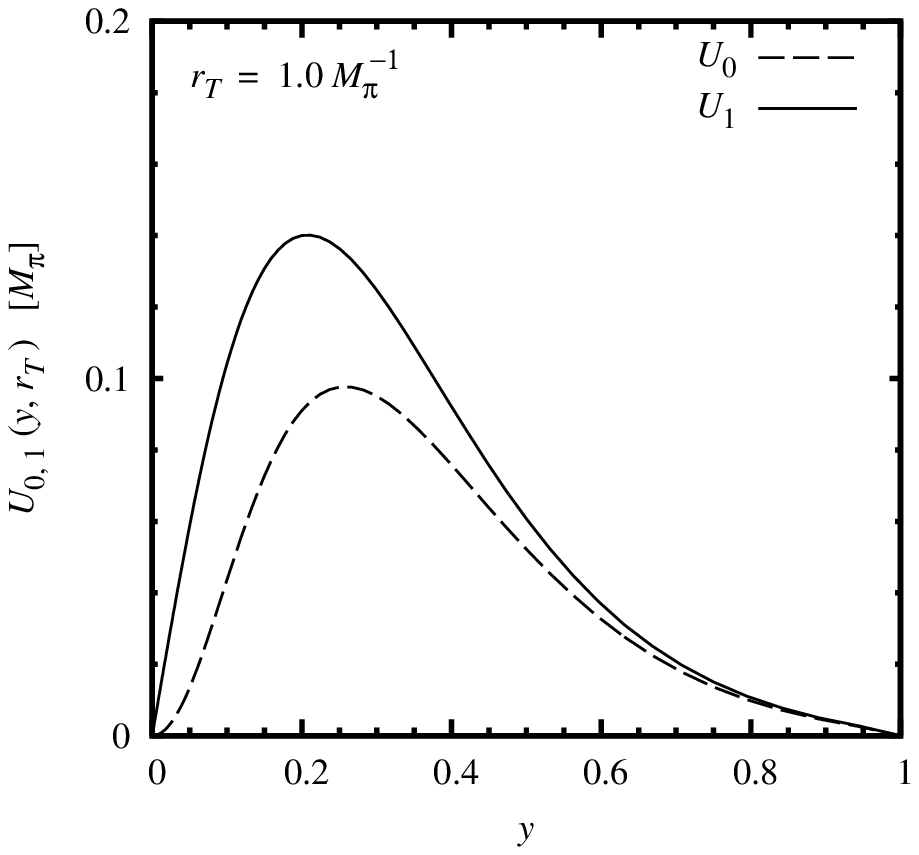}
&
\includegraphics[width=.48\textwidth]{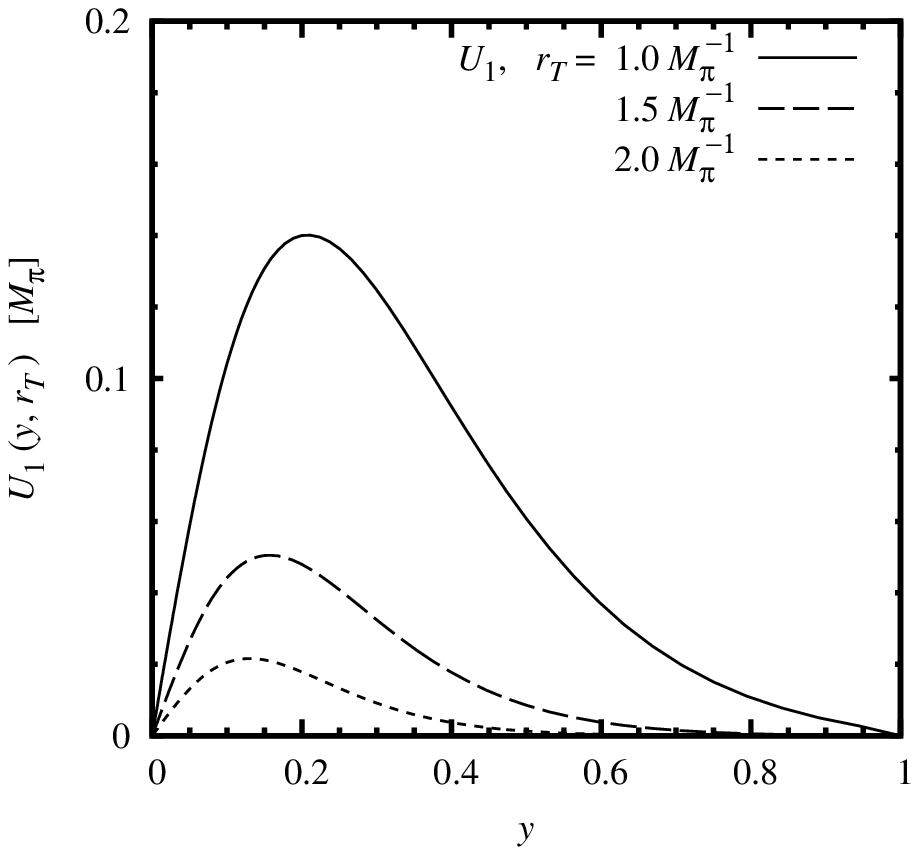}
\\[-2ex]
(a) & (b)
\end{tabular}
\caption[]{Peripheral chiral light--front wave function in coordinate 
representation. (a) Radial wave functions $U_0$ and $U_1$,
Eq.~(\ref{psi_0_1_def}) as functions of the pion momentum fraction $y$, 
at fixed transverse separation $r_T = 1.0 \, M_\pi^{-1}$.
(b) Radial wave functions $U_1$ as function of $y$ at several
transverse separations, $r_T = (1.0, 1.5, 2.0) \, M_\pi^{-1}$.}
\label{fig:wf_rt}
\end{figure}

Figure~\ref{fig:wf_rt} shows a plot of the peripheral radial wave
functions $U_{0, 1}(y, r_T)$ as functions of the pion momentum
fraction $y$. Plot (a) compares $U_0$ and $U_1$ at a fixed 
transverse separation. One sees that the $L^z = 0$ and $L^z = 1$
components become equal at $y \rightarrow 1$ (i.e., at values several 
times $M_\pi / M_N$), but show different power-like behavior
at $y \rightarrow 0$, as is already apparent from the analytic formulas
Eq.~(\ref{psi_0_1_def}). One also sees that the inequality 
Eq.~(\ref{psi_0_psi_1_inequality}) is satisfied. Plot (b) shows
the $L^z = 1$ wave function $U_1$ at several transverse separations.
One sees that values of $y \sim 1$ are strongly suppressed with
increasing transverse separation, and that the maximum of the wave 
function in $y$ shifts to smaller values, in accordance with general 
expectations. At $r_T = \textrm{several times}\; M_\pi^{-1}$ the wave
function -- and in particular the probabilities -- are strongly 
concentrated at pion momentum fractions $y = O(M_\pi/M_N)$,
and the parametric approximations are borne out by the numerical results.
%
% FIGURE
%
\section{Peripheral transverse densities}
\label{sec:densities}
\subsection{Overlap representation}
\label{subsec:overlap_densities}
We now want to express the peripheral transverse densities in the nucleon
in terms of the chiral light--front wave functions. For this we first need
to obtain explicit expressions for the invariant form factors in terms
of the spin components of the current matrix element Eq.(\ref{me_general}).
Taking the nucleon spin states in Eq.(\ref{me_general}) as light--front 
helicity states, cf.~Eq.~(\ref{spinor_boosted}), and choosing a frame 
where the momentum transfer has only transverse components,
cf.~Eq.~(\ref{Delta_transverse}), the matrix element of the plus
component of the current has the form 
\be
\langle N_2 | \, J^+ (0) \, | N_1 \rangle
&\equiv& 
\langle N(p^+, \bm{p}_{2T}, \sigma_2 ) |
\, J^+ (0) \, | N(p^+, \bm{p}_{1T}, \sigma_1 ) \rangle
\nonumber
\\[1ex]
&=& 
(2p^+) \left[ \delta(\sigma_2, \sigma_1) \; F_1 (-\bm{\Delta}_T^2) 
\; + \; 
i\left(\bm{\Delta}_T\times \bm{e}_z \right)\cdot
\bm{S}_T (\sigma_2, \sigma_1) \; \frac{F_2 (-\bm{\Delta}_T^2)}{M_N}
\right] .
\label{me_plus}
\ee
where $\bm{S}_T$ is defined as in Eq.~(\ref{S_vector}) in terms of the 
rest--frame 2--spinors describing the initial and final nucleon,
and $\bm{e}_z$ is the unit vector in $z$--direction.
The form factors $F_1$ and $F_2$ are then obtained from the 
diagonal and off-diagonal matrix elements as\footnote{The 
identification of the different spin components can be done
conveniently by writing both sides of Eq.~(\ref{me_plus})
as bilinear forms in the nucleon two--spinors, stripping off the
two-spinors, and treating the equation as a $2\times2$ matrix equation. 
The different components can then be projected out by taking 
appropriate traces of both sides.}
\be
\left. 
\begin{array}{l}
F_1 (-\bm{\Delta}_T^2)  \\[4ex] F_2 (-\bm{\Delta}_T^2) 
\end{array}
\right\}
&=& \frac{1}{2p^+} \; \sum_{\sigma_1\sigma_2}
\langle N_2 | \, J^+ (0) \, | N_1 \rangle
\left\{\begin{array}{c}
\displaystyle \frac{1}{2} \delta(\sigma_1, \sigma_2) 
\\[2ex]
\displaystyle 
\frac{2 M_N}{\bm{\Delta}_T^2} \;
(-i) (\bm{\Delta}_T\times \bm{e}_z)
\cdot \bm{S}_T(\sigma_1, \sigma_2)
\end{array}
\right\} .
\label{FF_matrix}
\ee
For the intermediate--nucleon part of the current matrix element we 
now substitute the overlap representation
in terms of the light--front wave functions in the transverse rest frame,
Eq.~(\ref{me_onshell_overlap_restframe}). Using the coordinate 
representation the overlap integral becomes diagonal in the transverse
relative coordinate $\bm{r}_T$ and takes the form
\be
\langle N_2 | \, J^+ (0) \, |N_1 \rangle^V_{\rm interm}
&=& 
\frac{(2 p^+)}{2\pi} \int\frac{dy}{y\bar y} \int d^2 r_T \;
e^{- i\bar y \bm{r}_T \bm{\Delta}_T}
\sum_{\sigma} 
\Phi^\ast (y, \bm{r}_T; \sigma , \sigma_2) \;
\Phi (y, \bm{r}_T; \sigma, \sigma_1) .
\label{me_overlap_restframe_coordinate}
\ee
Notice that the momentum transfer $\bm{\Delta}_T$ is Fourier--conjugate
not to $\bm{r}_T$ itself but to $\bar y \bm{r}_T$, which is a general
feature of light--front kinematics. It is now straightforward to 
evaluate the spin sums in Eq.~(\ref{FF_matrix}) and obtain the invariant 
form factors in terms of the $L^z = 0$ and $1$ components of the 
coordinate--space wave function Eq.~(\ref{psi_coordinate_decomposition}).
We immediately quote the results for the isovector transverse densities
$\rho_1^V$ and $\widetilde\rho_2^V$, Eq.~(\ref{rho_2_tilde_def}):
\be
\left. 
\begin{array}{l}
\rho_1^V(b) \\[3ex] \widetilde\rho_2^V(b) 
\end{array}
\right\}
&=& 
\frac{1}{2\pi} \int\frac{dy}{y\bar y^3}
\left\{
\begin{array}{c}
\displaystyle
[U_0(y, b/\bar{y})]^2 \; + \; 
[U_1(y, b/\bar{y})]^2
\\[3ex]
\displaystyle
- 2 \, 
U_0(y, b/\bar{y})\;
U_1(y, b/\bar{y})
\end{array}
\right\} .
\label{rho_overlap}
\ee

The form of Eq.~(\ref{rho_overlap}) is explained by the
spin structure of the transitions (see Fig.~\ref{fig:diag_wf}b). 
The light--front wave function 
has a nucleon helicity--conserving ($U_0$) and a helicity--flipping
component ($U_1$). The current matrix element with the same nucleon
helicity in the initial and final state requires the combination of
two helicity--conserving or two helicity--flipping wave functions
($U_0^2$ or $U_1^2$), whereas the matrix element with different
nucleon helicities in the initial and final state requires combination
of one helicity-conserving and one helicity-flipping wave function
($U_0 U_1$).

The transverse charge density $\rho_1^V(b)$ also receives a contribution
from the effective contact term in the current matrix element, 
Eq.~(\ref{me_contact}). This contribution cannot be represented as 
an overlap of $\pi N$ light--front wave functions and has to be 
added to Eq.~(\ref{rho_overlap}) as a separate term.
The exact form of this term and its interpretation are discussed
in Sec.~\ref{subsec:contact}. The numerical contribution of the 
contact term turns out to be very small at distances
$b = \textrm{few} \; M_\pi^{-1}$, so that the entire $\rho_1^V$ 
is to good approximation given by the wave function 
overlap Eq.~(\ref{rho_overlap}). We may therefore compare the 
properties of the densities $\rho_1^V$ and $\widetilde\rho_2^V$ 
on the basis of Eq.~(\ref{rho_overlap}) (the contact term is
absent in $\widetilde\rho_2^V$).
\subsection{Chiral order and inequality}
The overlap representation Eq.~(\ref{rho_overlap}) reveals several
interesting properties of the chiral component of the peripheral 
transverse densities. First, because the light--front helicity--conserving 
and --flipping wave functions appear in the same order of the chiral
expansion (their coefficients involve the same power of $M_\pi/M_N$),
cf.\ Eq.(\ref{psi_0_psi_1_ratio}), we conclude that the peripheral
densities $\rho_1^V$ and $\widetilde\rho_2^V$ are of the same order
in the chiral expansion,
\beq
\widetilde\rho_2^V (b) / \rho_1^V (b) \;\; = \;\; O(1) 
\hspace{2em} [b = O(M_\pi^{-1})].
\label{rho_2_tilde_rho_1_order}
\eeq
While this circumstance was noted earlier in the dispersive 
approach \cite{Granados:2013moa}, where it is encoded 
in the chiral order of the spectral functions of the form factors
near threshold, the overlap representation exhibits it more directly 
and provides a more immediate physical explanation. Notice that the
original transverse magnetization density $\rho_2^V$ is parametrically
larger than $\widetilde\rho_2^V$ by a power of $M_N/M_\pi$, because the
spatial derivative in Eq.~(\ref{rho_2_tilde_def}) ``counts'' as
$O(M_\pi)$ in the region $b = O(M_\pi^{-1})$, and thus
\beq
\rho_2^V (b) / \rho_1^V (b) \;\; = \;\; O(M_N/M_\pi) .
\eeq
The parametric equality of $\rho_1^V$ and $\widetilde\rho_2^V$, 
Eq.~(\ref{rho_2_tilde_rho_1_order}), allows for non-trivial dynamical
relations between the two densities and provides additional motivation
for working with $\widetilde\rho_2^V$ rather than $\rho_2^V$.

%
% FIGURE
%
\begin{figure}[t]
\parbox[c]{.48\textwidth}{
\includegraphics[width=.48\textwidth]{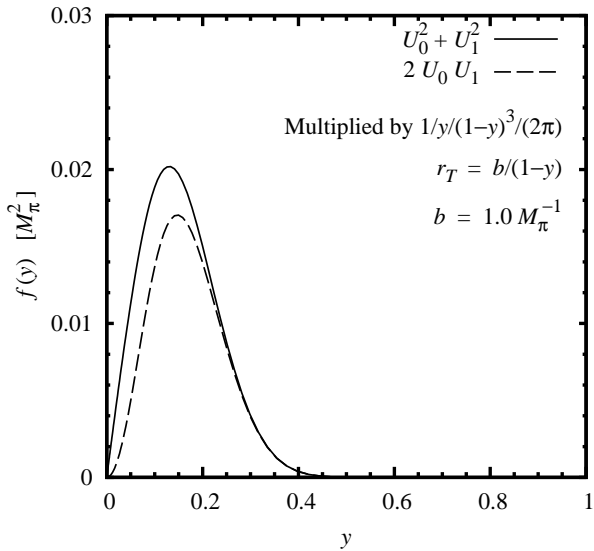}}
\hspace{.05\textwidth}
\parbox[c]{.4\textwidth}{
\caption[]{Integrands of the transverse densities $\rho_1^V$ and
$-\widetilde\rho_2^V$ in the wave function overlap representation
Eq.~(\ref{rho_overlap}), at a distance $b = 1.0 \, M_\pi^{-1}$.
The integrands are given in units of $M_\pi^{-2}$.}
\label{fig:wf_b}}
\end{figure}
Second, we observe an inequality between the peripheral transverse densities,
\beq
|\widetilde\rho_2^V(b)| \; \leq \; |\rho_1^V(b)| .
\label{inequality}
\eeq
It follows from the inequality obeyed by the quadratic forms in the 
integrands of the $y$--integral in Eq.~(\ref{rho_overlap}),
$U_0^2 + U_1^2 \geq 2 U_0 U_1$. While Eq.~(\ref{inequality}) 
was observed accidentally in the numerical calculations of 
Ref.~\cite{Granados:2013moa}, its mathematical proof and physical 
explanation become possible with the wave function overlap 
representation provided here. Recalling that $\rho_1^V(b)$ and 
$\widetilde\rho_2^V(b)$ represent the spin--independent and --dependent
components of the ``plus'' current at transverse position $b$ in a
nucleon localized at the origin and polarized along the $y$--direction, 
cf.~Eq.~(\ref{j_plus_rho}) and Fig.~\ref{fig:interpretation}, 
we see that the inequality Eq.~(\ref{inequality}) 
implies the positivity condition
\beq
\langle J^+ (\bm{b}) \rangle_{\text{\scriptsize localized}} \; \geq \; 
|\rho_1^V(b)| - |\widetilde\rho_2^V(b)| \; \geq \; 0 .
\label{positivity}
\eeq
This property appears natural when one realizes that the plus current at
$b = O(M_\pi^{-1})$ results from peripheral pions, and that the current 
carried by an on--shell pion is proportional to its 4--momentum,
$\langle \pi^+ (k) | \, J^+ \, | \pi^+ (k) \rangle = 2 k^+ > 0$.
Such a ``quantum--mechanical'' picture of the peripheral densities 
will be explored further in Sec.~\ref{subsec:mechanical}.
\subsection{Numerical evaluation}
\label{subsec:numerical}
We now want to use the overlap representation Eq.~(\ref{rho_overlap})
to study the numerical behavior of the transverse densities.
Figure~\ref{fig:wf_b} shows the integrands of $\rho_1(b)$ 
and $-\widetilde\rho_2(b)$ as functions of the
pion momentum fraction $y$ at a fixed transverse distance 
$b = 1.0\, M_\pi^{-1}$. One sees that the integrands are
concentrated around values $y \sim M_\pi / M_N$. Contributions
from $y > 0.5$ are very strongly suppressed because the wave function
is evaluated at separations $r_T = b/(1 - y)$ that are substantially
larger than $b$, and the wave function decays exponentially at large
$r_T$ with a range that itself decreases with increasing $y$.
One also sees that the integrands for the densities $\rho_1^V$
and $-\widetilde\rho_2^V$ are close to each other at large values 
of $y$ and differ only at $y \rightarrow 0$, such that they are numerically
close throughout the dominant region of integration. This follows from 
the similarity of the wave functions $U_0$ and $U_1$, 
cf.~Fig.~\ref{fig:wf_rt}a, and implies that $\rho_1^V(b) \approx
-\widetilde\rho_2^V(b)$ at distances $b = \textrm{few times} \, M_\pi^{-1}$.

\begin{figure}[t]
\parbox[c]{.48\textwidth}{
\includegraphics[width=.48\textwidth]{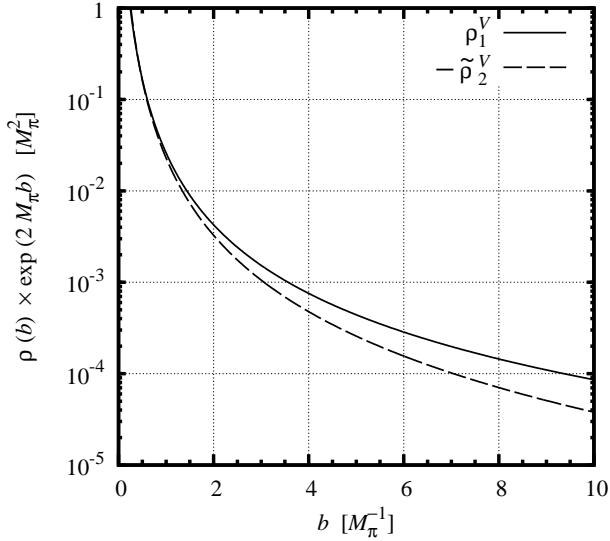}}
\hspace{.05\textwidth}
\parbox[c]{.4\textwidth}{
\caption[]{LO chiral component of the nucleon's 
isovector spin--independent current density $\rho_1^V(b)$ 
(solid line) and spin--dependent current density 
$-\widetilde \rho_2^V(b)$ (dashed line). 
The plot shows the densities
with the exponential factor $\exp (- 2 M_\pi b)$ extracted
[the functions plotted correspond to the pre-exponential factor
in Eq.~(\ref{large_b_general})].}
\label{fig:rho12t_scaled}
}
\end{figure}

The transverse densities obtained by performing the $y$--integral 
in Eq.~(\ref{rho_overlap}) are shown in Fig.~\ref{fig:rho12t_scaled}. 
One sees that $\widetilde\rho_2^V < 0$,
and that the absolute value of the spin--dependent density is smaller 
than the spin--independent one, $ |\widetilde\rho_2^V| < \rho_1^V$,
as required by Eq.~(\ref{inequality}). The inequality is almost saturated
at distances $b \sim 1 \, M_\pi^{-1}$, as suggested by the
integrands shown in Figure~\ref{fig:wf_b}, but at larger distances
$|\widetilde\rho_2^V|$ becomes significantly smaller than $\rho_1^V$.
We note that the numerical densities obtained from the wave function
overlap representation Eq.~(\ref{rho_overlap}) exactly reproduce
those calculated in the dispersive approach of Ref.~\cite{Granados:2013moa},
which provides a test of the calculational procedures.\footnote{In 
Fig.8 of Ref.~\cite{Granados:2013moa} the function represented by the 
dashed line is $-\widetilde\rho_2(b)$, not $\widetilde\rho_2(b)$
(the plot is labeled incorrectly). The dispersive calculation gives 
$\widetilde\rho_2(b) < 0$, as does the wave function representation
of the present work.}
%
% FIGURE
%
\subsection{Transverse polarization}
\label{subsec:transverse_polarization}
Further insight into the peripheral transverse densities can be gained
by considering the case of transversely polarized nucleon states.
Transverse polarization naturally explains the 
similarity of the spin--independent and --dependent densities, $\rho_1^V(b)$ 
and $\widetilde\rho_2^V(b)$, at distances $b \sim \textrm{few} \, M_\pi^{-1}$
(cf.~Fig.~\ref{fig:rho12t_scaled}) and enables a simple quantum--mechanical
interpretation in the nucleon rest frame.

Transversely polarized nucleon states in the light--front formulation
are obtained by preparing a transversely polarized state (say, in the 
$y$--direction) in the rest frame and performing the longitudinal and 
transverse boosts to the desired light--front momentum 
(cf.~Sec.~\ref{subsec:spin}). With the general formula 
Eq.~(\ref{spinor_boosted}) this is accomplished simply by choosing 
the rest--frame 2--spinors as eigenspinors of the $y$--spin operator
$S^y = \sigma^y/2$,
\beq
\chi_{\rm tr} (\tau = 1/2) \; = \; \frac{1}{\sqrt{2}}
\left( \begin{array}{c} 1 \\ i \end{array}\right),
\hspace{2em}
\chi_{\rm tr} (\tau = -1/2) \; = \; \frac{1}{\sqrt{2}}
\left( \begin{array}{c} i \\ 1 \end{array}\right)
\label{chi_along_y}
\eeq
(here and in the following we use $\tau = \pm 1/2$ to denote the $y$--spin 
eigenvalues). The vertex function calculated with the transversely 
polarized 4--spinors obtained in this way is of the same form as
Eq.~(\ref{vertex_explicit}); only the structures $S^z$ and 
$\bm{S}_T = (S^x, S^y)$ are replaced by the contraction of 
the spin operators with the transversely polarized spinors
\be
S^i_{\rm tr} (\tau, \tau_1) &\equiv& \chi^\dagger_{\rm tr} (\tau)
({\textstyle\frac{1}{2}} \sigma^i ) \chi_{\rm tr} (\tau_1) 
\hspace{2em} (i = x, y, z).
\label{S_vector_y}
\ee
Notice that now the $y$--component is diagonal,
\beq
S_{\rm tr}^y (\tau, \tau_1) 
\;\; = \;\; \tau \; \delta(\tau, \tau_1) ,
\eeq
while the $z$ and $x$ components have off--diagonal terms. We define
the light--front wave function for the $N \rightarrow \pi N$ transition
for nucleon states characterized by their transverse spin projections
$\tau_1$ and $\tau$, in complete analogy to 
Eqs.~(\ref{psi_restframe})--(\ref{vertex_restframe}),
\be
\Psi_{\rm tr} (y, \widetilde{\bm{k}}_T; \tau, \tau_1)
&\equiv& 
\frac{\Gamma_{\rm tr} (y, \widetilde{\bm{k}}_T; \tau, \tau_1)}
{\Delta\mathcal{M}^2(y, \widetilde{\bm{k}}_T)} ,
\label{psi_restframe_y}
\\[1ex]
\Gamma_{\rm tr} (y, \widetilde{\bm{k}}_T; \tau, \tau_1)
&=& \frac{2i g_A M_N}{F_\pi \sqrt{\bar{y}}}
\left[ y M_N \, S_{\rm tr}^z (\tau, \tau_1) + 
\widetilde{\bm{k}}_T \cdot \bm{S}_{\rm tr, T} (\tau, \tau_1) 
\right] .
\label{vertex_restframe_y}
\ee
Similar expressions are obtained for the $\pi N \rightarrow N$ with 
transverse spin projections $\tau$ and $\tau_2$.

The coordinate--space wave functions for transverse nucleon polarization 
are introduced through Eq.~(\ref{psi_coordinate}) in the same way as 
for longitudinal polarization and denoted by $\Phi_{\rm tr}$. 
The general decomposition 
Eq.~(\ref{psi_coordinate_decomposition}) applies to the Fourier transform
of the transversely polarized wave function as well, as it relies only on 
the functional dependence of the momentum--space wave function 
on $\widetilde{\bm{k}}_T$, not on the specific form of the spin structures.
Using the algebraic relation between the spin structures for $y$-- and
$z$--polarization it is straightforward to express the transversely
polarized coordinate--space wave function in terms of the longitudinally
polarized radial wave functions $U_0$ and $U_1$:
\be
\Phi_{\rm tr} (y, \, \bm{r}_T, \, 
\tau = +1/2, \, \tau_1 = +1/2) &=&  \phantom{-}\sin\alpha \, U_1 ,
\label{psi_ypol_plus_plus}
\\[1ex]
\Phi_{\rm tr} (y, \, \bm{r}_T, \, 
\tau = -1/2, \, \tau_1 = -1/2) &=& -\sin\alpha \, U_1 ,
\label{psi_ypol_minus_minus}
\\[1ex]
\Phi_{\rm tr} (y, \, \bm{r}_T, \, 
\tau = +1/2, \, \tau_1 = -1/2) &=& 
\phantom{-} U_0 \; + \; \cos\alpha \, U_1 ,
\label{psi_ypol_plus_minus}
\\[1ex]
\Phi_{\rm tr} (y, \, \bm{r}_T, \, 
\tau = -1/2, \, \tau_1 = +1/2) &=& 
- U_0 \; + \; \cos\alpha \, U_1 ,
\label{psi_ypol_minus_plus}
\\[1ex]
\bm{r}_T \; = \; (r_T \cos\alpha, \, r_T \sin\alpha), &&
U_{0,1} \; \equiv \; U_{0,1}(y, r_T) .
\nonumber
\ee
With the radial wave function $U_0$ and $U_1$ given by
the explicit expressions of Eq.~(\ref{psi_0_1_def}), these relations 
completely determine the transversely polarized coordinate--space 
wave function. 

In the transversely polarized representation the rotational symmetry 
around the $z$--axis is encoded in relations between the transverse spin 
components of the wave function. In the explicit formulas
Eqs.~(\ref{psi_ypol_plus_plus})--(\ref{psi_ypol_minus_plus}) 
these relations manifest themselves in that the four transverse spin 
components are expressed in terms of only two independent 
radial functions. One sees that the components satisfy
\be
\Phi_{\rm tr} (y, \, \bm{r}_T, \, +, \, +) 
&=& -\Phi_{\rm tr} (y, \, \phantom{-}\bm{r}_T, \, -, \, -) ,
\label{psi_ypol_relation_nonflip}
\\[1ex]
\Phi_{\rm tr} (y, \, \bm{r}_T, \, +, \, -) 
&=& -\Phi_{\rm tr} (y, \, -\bm{r}_T, \, -, \, +) .
\label{psi_ypol_relation_plus}
\ee
The relation Eq.~(\ref{psi_ypol_relation_plus}) between the transverse
spin--flip components has a simple physical interpretation in the nucleon
rest frame. In the wave function with initial nucleon transverse
spin $\tau_1 = +1/2$ and intermediate nucleon spin $\tau = -1/2$, 
the pion in the intermediate state has orbital angular momentum 
$L = 1$ with projection $L^y = +1$ on the $y$--axis. 
Likewise, in the wave function 
with $\tau_1 = -1/2$ and $\tau = +1/2$, the pion has $L^y = -1$. 
The two components thus differ only in that the pion rotates in the
opposite sense around the $y$--axis, and one can be turned into the other
by inverting the direction of the $x$--axis, i.e., replacing
$\cos\alpha \rightarrow -\cos\alpha$ in Eqs.~(\ref{psi_ypol_plus_minus})
and (\ref{psi_ypol_minus_plus}).

The connection between the transversely and longitudinally polarized 
light--front wave 
functions generally depends on the angle of the transverse coordinate 
vector $\bm{r}_T$, as required by rotational invariance. A particularly
simple connection is obtained at points on the negative or positive
$x$--axis, where $\sin\alpha = 0$ and $\cos\alpha = -1$ or $+1$.
Considering the transverse spin--flip wave function on the negative and 
positive $x$--axis (``left'' and ''right'' when looking at the nucleon 
in the $z$--direction from $+\infty$, see Fig.~\ref{fig:interpretation}), 
and introducing the short--hand
notation
\be
U_{\rm left} (y, \, r_T)
&\equiv& \Phi_{\rm tr} (y, \, - r_T \bm{e}_x, \, \tau = -1/2, \, 
\tau_1 = +1/2) ,
\label{psi_left}
\\[1ex]
U_{\rm right} (y, \, r_T)
&\equiv& \Phi_{\rm tr} (y, \,  +r_T \bm{e}_x, \, \tau = -1/2, \, 
\tau_1 = +1/2) ,
\label{psi_right}
\ee
we obtain from Eq.~(\ref{psi_ypol_minus_plus})
\be
U_{\rm left} (y, \, r_T)
&=& - U_0(y, \, r_T) \, - \, U_1(y, \, r_T) ,
\label{psi_left_connection}
\\[1ex]
U_{\rm right} (y, \, r_T)
&=& - U_0(y, \, r_T) \, + \, U_1(y, \, r_T) .
\label{psi_right_connection}
\ee
Note that the functions $U_{\rm left}$ and $U_{\rm right}$ are
derived from a \textit{single} $y$--spin component of the wave function
but refer to a specific spatial direction.

We can now derive a simple representation of the transverse densities 
in terms of the transversely polarized light--front wave function. 
Using the overlap representation Eq.~(\ref{rho_overlap}) in terms of the 
the longitudinally polarized wave functions, and substituting them by 
the ``left'' and ``right'' transverse spin--flip wave functions according 
to Eqs.~(\ref{psi_left_connection}) and (\ref{psi_right_connection}), 
we obtain\footnote{For simplicity we derive the representation 
Eq.~(\ref{rho_overlap_left_right}) from Eq.~(\ref{rho_overlap}), 
using the relation between the longitudinally and transversely polarized
wave functions at the special points $b = \pm b \bm{e}_x$. 
Equation~(\ref{rho_overlap_left_right}) could equivalently be derived 
by converting the original current matrix element Eq.~(\ref{me_plus}) 
to the transverse spin representation and repeating the steps of 
Sec.~\ref{subsec:overlap_densities}. In the latter approach one could
choose any orientation of the vector $\bm{b}$; the rotational invariance
of the densities would be guaranteed by the conditions 
Eqs.~(\ref{psi_ypol_relation_nonflip}) and (\ref{psi_ypol_relation_plus}).}
\be
\left. 
\begin{array}{l}
\rho_1^V (b) \\[3ex] \widetilde\rho_2^V (b) 
\end{array}
\right\}
&=& 
\frac{1}{4\pi} \int\frac{dy}{y\bar y^3}
\left\{
\begin{array}{c}
\displaystyle \phantom{-}
[U_{\rm left}(y,b/\bar{y})]^2 \; + \; 
[U_{\rm right}(y,b/\bar{y})]^2
\\[3ex]
\displaystyle
- [U_{\rm left}(y,b/\bar{y})]^2 \; + \; 
[U_{\rm right}(y,b/\bar{y})]^2
\end{array}
\right\} .
\label{rho_overlap_left_right}
\ee
This result can be explained easily by referring to the interpretation 
of the transverse densities as current matrix elements in a nucleon
state localized in transverse space, cf.\
Sec.~\ref{subsec:transverse_densities},
Fig.~\ref{fig:interpretation}, and Ref.~\cite{Granados:2013moa}.
According to Eq.~(\ref{j_plus_rho}) in a nucleon state with $y$--spin 
projection $+1/2$ (i.e., $\tau_1 = \tau_2 = +1/2$) the dependence of 
the matrix element on the coordinate $\bm{b}$ is of the form
\be
\langle J^+ (\bm{b}) \rangle^V_{\text{\scriptsize localized}}
&=& (...) \; \left[
\rho_1^V (b) \;\; + \;\; \cos\phi \, \widetilde\rho_2^V (b) \right] ,
\ee
where $\phi$ is the angle of $\bm{b}$ relative to the $x$--axis.
The densities are functions of $b = |\bm{b}|$, and the angular dependence 
is given entirely by $\cos\phi$. By choosing the direction of $\bm{b}$
along the positive and negative $x$--axis ($\cos\phi = \pm 1$), the
individual densities can be expressed as
\be
\langle J^+ (b\bm{e}_x) \; + \; 
J^+ (-b\bm{e}_x) \rangle^V_{\text{\scriptsize localized}}
&=& 
(...) \; \rho_1^V (b) ,
\\[1ex]
\langle J^+ (b\bm{e}_x) \; - \; 
J^+ (-b\bm{e}_x) \rangle^V_{\text{\scriptsize localized}}
&=& 
(...) \; \widetilde\rho_2^V (b) .
\ee
This structure is exactly analogous to Eq.~(\ref{rho_overlap_left_right}),
with the integral over $|U_{\rm left}(y, b/\bar y)|^2$ 
and $|U_{\rm left}(y, b/\bar y)|^2$
representing the current densities at $b = \pm b \bm{e}_x$, respectively.

%
% FIGURE
%
\begin{figure}[t]
\parbox[c]{.48\textwidth}{
\includegraphics[width=.48\textwidth]{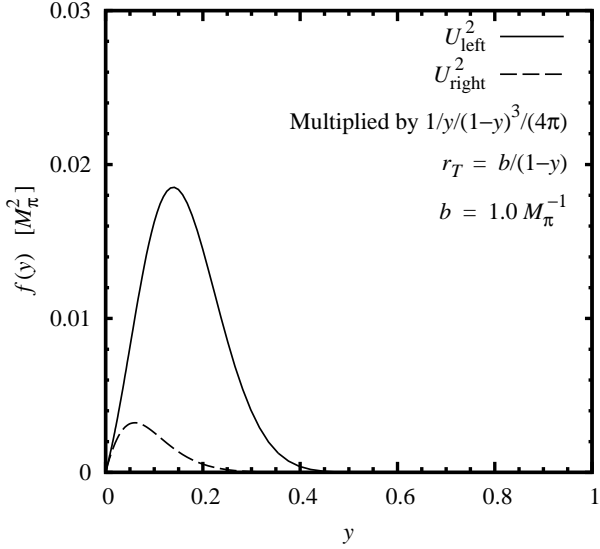}}
\hspace{.05\textwidth}
\parbox[c]{.4\textwidth}{
\caption[]{Integrands of the transverse densities $\rho_1^V$ and
$\widetilde\rho_2^V$ in the transversely polarized wave function 
overlap representation Eq.~(\ref{rho_overlap_left_right}), at a 
distance $b = 1.0 \, M_\pi^{-1}$. Solid line: 
$[U_{\rm left}(y, b/\bar y)]^2$; dashed line: 
$[U_{\rm right}(y, b/\bar y)]^2$.
The functions are plotted including the factor 
$1/(4 \pi y\bar y^3)$ and are given in 
units of $M_\pi^{-2}$.}
\label{fig:wf_transv_b}}
\end{figure}

The explicit form of the ``left'' and ''right'' transverse spin--flip
wave function, Eqs.~(\ref{psi_left}) and (\ref{psi_right}),
is readily obtained from Eqs.~(\ref{psi_left_connection}), 
(\ref{psi_right_connection}) and (\ref{psi_0_1_def}):
\be
\left.
\begin{array}{r}
 U_{\rm left}(y,r_T) 
\\[2ex]
 U_{\rm right}(y,r_T)
\end{array}
\right\}
&=& -\frac{g_A M_N \, y \sqrt{\bar{y}}}{2\pi F_\pi}
\left\{
\begin{array}{r}
[y M_N \, K_0(M_T r_T)  + M_T K_1(M_T r_T)]
\\[2ex]
[y M_N \, K_0(M_T r_T)  - M_T K_1(M_T r_T)]
\end{array}
\right\} 
\label{psi_left_right_explicit}
\\[2ex]
&\sim& - \frac{g_A M_N \, y \sqrt{\bar y}}{2 \sqrt{2\pi} F_\pi}
\left\{
\begin{array}{r}
(y M_N + M_T)
\\[2ex]
(y M_N - M_T)
\end{array}
\right\}
\;\frac{e^{-M_T r_T}}{\sqrt{M_T r_T}} 
\hspace{4em} (r_T M_T \gg 1) .
\label{psi_left_right_asymptotic}
\ee
The last expression is obtained with the asymptotic form of the modified 
Bessel functions, Eq.~(\ref{K01_asymptotic}), and applies at $M_T r_T \gg 1$.
Parametrically the two functions are of the same order in $M_\pi / M_N$
in the region $y = O(M_\pi / M_N)$ and $r_T = O(M_\pi^{-1})$. Numerically
one observes that
\beq
|U_{\rm left}(y,r_T)| 
\;\; \gg \;\; 
|U_{\rm right}(y,r_T)| 
\hspace{2em} (y \sim \textrm{few times} \; M_\pi/M_N) ,
\eeq
because $M_T (y) \approx y M_N$ in this region of $y$.
This is illustrated by Fig.~\ref{fig:wf_transv_b}, which shows
the contributions of $|U_{\rm left}|^2$ and $|U_{\rm right}|^2$ 
to the integrands of the transverse
densities in Eq.~(\ref{rho_overlap_left_right}). The strong suppression
of the ``right'' compared to the ``left'' wave function is the reason
for the similarity of the densities $\rho_1^V(b)$ and $\widetilde\rho_2^V (b)$
at $b = O(M_\pi^{-1})$.
\subsection{Quantum--mechanical picture}
\label{subsec:mechanical}
We can interpret our findings in a simple quantum--mechanical picture 
of peripheral transverse nucleon structure in the rest frame.
For this purpose we imagine that the $N \rightarrow \pi N$ transition 
takes place in ordinary time, and that the wave function has the usual 
3--dimensional rotational symmetry. For a non-relativistic system
there would be a direct correspondence between the equal--time 
and the light--front wave functions; see e.g.\ Ref.~\cite{Frankfurt:1981mk}. 
The chiral $\pi N$ system is essentially relativistic, 
$k = O(M_\pi)$, and one should not expect a similar connection
between the wave functions here. Nevertheless the intuitive 
quantum--mechanical picture of the chiral process explains all the 
essential features of the peripheral transverse densities. In any case 
its content is backed up by light--front wave function formulas, 
which are exact also in the relativistic case.

\begin{figure}[t]
\begin{center}
\includegraphics[width=.6\textwidth]{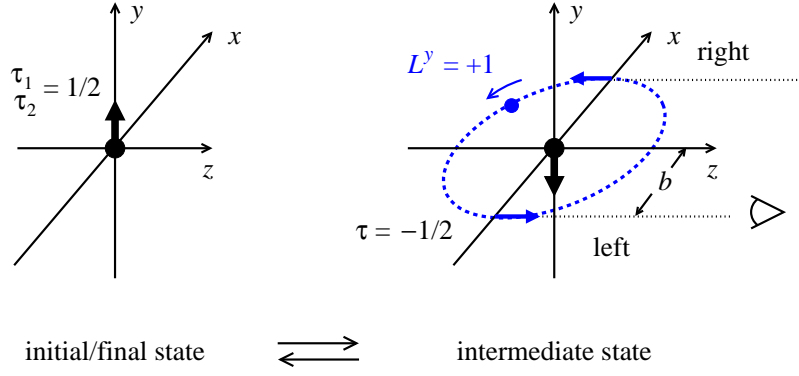}
\end{center}
\caption[]{Quantum--mechanical picture of peripheral transverse 
densities in LO chiral EFT. The bare nucleon with $y$--spin projection
$\tau_1 = +1/2$ in the rest frame (left) makes a transition to a
pion--nucleon state with $\tau = -1/2$ and $L^y = +1$ (right),
and back to a bare state with $\tau_2 = +1/2$. 
The peripheral transverse densities 
$\rho_1^V (b)$ and $\widetilde\rho_2^V (b)$ are the left--right average
and left--right asymmetry of the plus current 
(viewed from $z = +\infty$) at the positions $y = 0$ and $x = \mp b$.
\label{fig:mech}}
\end{figure}

Consider a nucleon in the rest frame, in a spin state polarized in 
the positive $y$--direction ($S^y = +1/2$). In the interaction 
picture implied by chiral EFT, we may think of this 
physical nucleon as a pointlike bare nucleon coupled to soft pions, 
described by a wave function. The LO contribution to the peripheral charge 
and current densities in the proton isospin state arises from the component 
with a single peripheral positively charged pion. This component corresponds 
to the chiral process where the initial (bare) proton makes a transition 
to a state with a (bare) neutron and a peripheral positive pion,
and back to the final bare nucleon (see Fig~\ref{fig:mech}). 
The $y$--spin projection quantum numbers of the initial/final nucleon 
state are $\tau_1 = \tau_2 = +1/2$, and that of the intermediate nucleon 
state is denoted by $\tau$. 
Because of parity conservation the wave function of the $\pi^+ n$ 
system has orbital angular momentum $L = 1$. 
Conservation of the total angular momentum projection on the $y$--axis
allows only the states with $L^y = 0$ and $\tau = +1/2$ 
(spin--conserving), and with $L^y = +1$ and $\tau = -1/2$ (spin--flip). 
We are interested in the densities in the $x$--$z$ plane ($y = 0$), 
to which the state with $L^y = 0$ cannot contribute, as its wave 
function vanishes in the direction perpendicular to the $y$--axis, 
$P_1(\cos\theta) = 0$ for $\theta = \pi/2$ ($P_1$ is the Legendre 
polynomial of degree 1). This leaves the state with $L^y = +1$ and 
$\tau = -1/2$ as the only contribution to the densities in question 
It explains why in the light--front formulation 
we were able to express the transverse densities completely in terms of 
the transverse spin--flip wave function, cf.~Eqs.~(\ref{psi_left})
and (\ref{psi_right}).
%
% FIGURE
%

The peripheral densities in the $x$--$z$ plane thus arise from 
configurations in which the positive pion ``orbits'' around the neutron
with angular momentum $L^y = +1$. Since the pion itself has no spin, 
the current it produces is the convection current caused by the
orbital motion of the charge. Because the chiral EFT interactions
between the pion and the nucleon have a short range $\ll M_\pi^{-1}$,
the peripheral pion can be regarded as free while the current is
measured. The 4--vector current carried by a free pion with 
momentum $\bm{k}$ and charge $+1$ is $J^\mu = 2 k^\mu$,
where $k^0 \equiv E_\pi = \sqrt{|\bm{k}|^2 + M_\pi^2}$, so that
its plus component is positive for all momenta,
$J^+ = 2k^+ = 2(E_\pi + k^z) > 0$. This explains the positivity 
property of the current density, Eq.~(\ref{positivity}).

Now according to Sec.~\ref{subsec:transverse_densities} and 
Eq.~(\ref{j_plus_rho}) the transverse densities $\rho_1^V$ and 
$\widetilde\rho_2^V$ are the left--right average and left--right asymmetry 
of the $J^+$ current density produced by the peripheral pion.
It is obvious that a positively charged pion orbiting with $L^y = +1$ 
produces a larger $J^+$ density on the left (where it moves in the positive
$z$ direction) than on the right (where it moves in the negative
$z$ direction). This explains why $\widetilde\rho_2^V(b) < 0$ (see
Fig.~\ref{fig:rho12t_scaled}).
The magnitude of the asymmetry is determined by the 
effective pion momenta and the relativistic effects implied by
the projection on fixed light--front time. If the motion of the
pion were non-relativistic, with characteristic velocity 
$v = k / M_\pi \ll 1$, the plus momentum carried by the pion
would be $k^+ = M_\pi [1 + O(v)]$; i.e., it would be dominated by the
pion mass and independent of the direction of the pion momentum.
Since furthermore the probability to find a pion would be the same on 
the left and on the right side of the $x$ axis (because of rotational
symmetry around the $y$--axis) the ratio of left and right current 
densities at the same distance $b = |\bm{b}|$ would be 
\beq
\frac{\langle J^+(-b\bm{e}_x) \rangle_{\text{\scriptsize localized}}}
{\langle J^+(+b\bm{e}_x) \rangle_{\text{\scriptsize localized}}}
\;\; = \;\; 1 + O(v).
\label{ratio_nonrel}
\eeq
The light--front wave function representation of 
Sec.~\ref{subsec:transverse_polarization} shows that the 
asymmetry is much larger than the non-relativistic 
estimate Eq.~(\ref{ratio_nonrel}),
\beq
\frac{\langle J^+(-b\bm{e}_x) \rangle_{\text{\scriptsize localized}}}
{\langle J^+(+b\bm{e}_x) \rangle_{\text{\scriptsize localized}}}
\;\; = \;\; 
\frac{\displaystyle \int\frac{dy}{y\bar y^3}
\left| U_{\rm left}(y,b/\bar{y}) \right|^2}
{\displaystyle \int\frac{dy}{y\bar y^3}
\left| U_{\rm right}(y,b/\bar{y}) \right|^2}
\;\; \gg \;\; 1.
\eeq
The numerical value of the ratio is $\sim 9$ at 
$b = 1\, M_\pi^{-1}$ and $\sim 4$ at $b = 5\, M_\pi^{-1}$
(see Fig.~\ref{fig:wf_transv_b}). This highlights the essentially 
relativistic nature of the motion of pions in chiral dynamics.
The power of the light--front formulation is that it permits 
a first--quantized representation even of such 
essentially relativistic systems.
\subsection{Contact term}
\label{subsec:contact}
\begin{figure}[t]
\parbox[c]{.48\textwidth}{
\includegraphics[width=.48\textwidth]{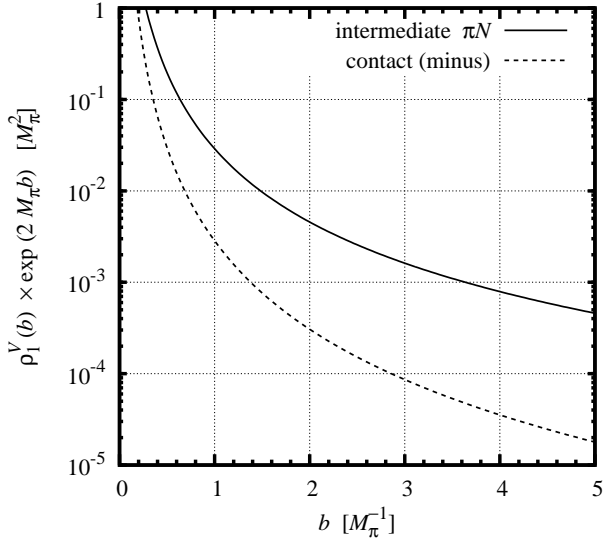}}
\hspace{.05\textwidth}
\parbox[c]{.4\textwidth}{
\caption[]{LO chiral component of the nucleon's 
isovector spin--independent current density $\rho_1^V(b)$.
Solid line: Contribution from intermediate $\pi N$ states,
Eqs.~(\ref{me_onshell}) and (\ref{rho_overlap}).
Dotted line: Effective contact term, Eq.~(\ref{me_contact}),
plotted with opposite sign (the actual contribution is negative). 
The plot shows the contributions to the density with the exponential 
factor $\exp (- 2 M_\pi b)$ extracted
[the functions plotted correspond to the pre-exponential factor
in Eq.~(\ref{large_b_general})].}
\label{fig:contact}
}
\end{figure}

The transverse charge density $\rho_1^V(b)$ also receives a contribution
from the effective contact term in the current matrix element, 
Eq.~(\ref{me_contact}). This contribution cannot be represented as 
an overlap of $\pi N$ light--front wave functions and needs to be 
added separately to that from the intermediate $\pi N$ state,
Eq.~(\ref{rho_overlap}),
\be
\rho_1^V(b) &=& \rho_1^V(b)_{\rm interm} \;\; 
[\textrm{from Eq.~(\ref{rho_overlap})}] \;\;\; 
+ \; \rho_1^V (b)_{\rm contact} .
\label{rho_intermediate_contact}
\ee
It is readily computed by evaluating the Feynman
integral Eq.~(\ref{me_contact}) as a four--dimensional integral, 
using the fact that it depends only on the momentum transfer $\Delta^\mu$
as external 4--vector. One obtains \cite{Granados:2013moa}
\be
\rho_1^V(b)_{\rm contact} &=& \frac{(1 - g_A^2) M_\pi^4}{192 \pi^3 F_\pi^2}
\left\{ [K_2(M_\pi b)]^2 - 4 [K_1(M_\pi b)]^2 + 3 [K_0(M_\pi b)]^2
\right\} .
\label{rho_contact}
\ee
The contact term density is negative because $1 - g_A^2 < 0$. 
It is of the same parametric order as the one from intermediate 
$\pi N$ states, as can be seen by comparing Eq.~(\ref{rho_contact}) 
with Eqs.~(\ref{rho_overlap}) and (\ref{psi_0_1_def}) and noting
that at distances $b = O(M_\pi^{-1})$ the integral is dominated by 
pion momentum fractions $y = O(M_\pi/M_N)$. The numerical
contribution of the contact term density turns out to be very small 
in the region of interest, $b = \textrm{few} \; M_\pi^{-1}$, ranging 
from about $-10\%$ of the intermediate $\pi N$ contributions at 
$b = 1 \, M_\pi^{-1}$ to $-4\%$ at $b = 5 \, M_\pi^{-1}$ (see 
Fig.~\ref{fig:contact}). Thus the entire $\rho_1^V$ is to good 
approximation given by the wave function overlap Eq.~(\ref{rho_overlap}),
which justifies our earlier comparison of the properties of $\rho_1^V$ and
$\widetilde\rho_2^V$ on the basis of Eq.~(\ref{rho_overlap}) (the contact 
term is absent in $\widetilde\rho_2^V$).
%
% FIGURE
%

Some comments are in order regarding the interpretation of the 
contact term in the context of the light--front description. 
In the intermediate $\pi N$ contribution to the current matrix
element the typical energy denominators are 
$\Delta\mathcal{M}^2(y, \bm{k}_T, \bm{p}_{1T, \, 2T}) = O(M_\pi M_N)$, 
cf.\ Eqs.~(\ref{virtuality_invariant_mass_1}) and
(\ref{virtuality_invariant_mass_2}), and the typical pion 
light--front energies in the nucleon rest frame are
$k_{1, 2}^- = O(M_\pi)$. These are configurations that exist for
a large light--front time interval $\Delta x^- = O(M_\pi^{-1})$ 
and can be regarded as particle states in the chiral EFT.
The contact term describes contributions to the peripheral
density from intermediate states with invariant mass differences
that are not chirally small, i.e., that do not vanish in the
limit $M_\pi \rightarrow 0$. These states lie ``outside'' the 
chiral EFT, and their contribution is represented by a local operator. 

The appearance of the combination $1 - g_A^2$ in the coefficient of 
the local operator is natural \cite{Granados:2013moa}. For a nucleon 
without internal structure one would have $g_A = 1$, and the 
effective contact term would be absent. The deviation of $g_A$ from
unity is the result of the ``compositeness'' of the nucleon, which in
turn is related to the presence of inelastic states in $\pi N$ scattering. 
The combination $1 - g_A^2$ can thus be regarded as the effect of 
non-chiral intermediate states in the $\pi N$ scattering amplitude,
in agreement with the above interpretation.\footnote{The 
connection between $g_A^2 - 1$ and inelastic states in
$\pi N$ scattering is expressed in general terms by the Adler--Weisberger 
current algebra sum rule \cite{Adler:1968hc,Weisberger:1966ip}.}
Further properties of the contact term, such as its relation to the
form of the $\pi NN$ coupling (pseudoscalar, axial vector) are
discussed in Ref.~\cite{Granados:2013moa}.

The contact term formally corresponds to a contribution of light--front 
zero modes (vanishing pion plus momentum) to the current matrix element. 
An advantage of our approach, starting from invariant integrals,
is that it allows us to identify and calculate these contributions
in a straightforward manner. The zero mode contributions could also
be calculated in the time--ordered formulation of chiral EFT, 
by considering the current matrix in a frame where the current 
transfers plus momentum, $\Delta^+ \neq 0$, and taking the limit 
$\Delta^+ \rightarrow 0$ at the end of the calculation.
\section{Chiral generalized parton distributions}
\label{sec:generalized}
\subsection{Peripheral pion distribution}
In QCD the transverse structure of the nucleon is expressed in terms of
coordinate--dependent parton densities, which are defined as the Fourier 
transforms of the generalized parton distributions and describe the 
density of partons (quarks, antiquarks, gluons) with a given 
light--front plus momentum fraction $x$ at transverse position $\bm{b}$.
In this context the transverse charge and current densities are 
obtained as integrals of the transverse densities of charged partons 
(quarks minus antiquarks) over $x$. Chiral dynamics governs not only 
the peripheral charge and current densities but also the $x$--dependent 
distributions of partons at $b \equiv |\bm{b}| = O(M_\pi^{-1})$ 
and $x = O(M_\pi/M_N)$. Detailed studies of the peripheral parton
densities due to chiral dynamics have been performed in 
Refs.~\cite{Strikman:2003gz,Strikman:2009bd}. Here we show 
that the $x$--integral of these peripheral parton densities reproduces 
the transverse charge and current densities calculated in chiral EFT.
We also express the plus momentum distribution of peripheral pions 
in the nucleon in terms of the $\pi N$ light--front wave functions
introduced in Sec.~\ref{sec:current_matrix_element}.

The basic object in the study of peripheral partonic structure is the
light--front plus momentum distribution of soft pions in the nucleon.
Following Refs.~\cite{Strikman:2003gz,Strikman:2009bd} we define the 
GPDs of soft pions in the nucleon in terms of the matrix element 
of the bilinear light--ray operator in the pion field
\be
&& p^+ \int\limits_{-\infty}^\infty
\frac{d\xi^-}{2\pi} \; e^{i y p^+ \xi^-/2} \; \langle N(p_2, \sigma_2) | 
\; \left[ \sum_{ab} \; \epsilon^{3ab} \; \pi^a (-\xi/2) 
\; \stackrel{\leftrightarrow}{\partial}{}^{\!\! +}
\pi^b (\xi/2) \; |_{\xi^+ = 0, \, \bm{\xi}_T = 0} \right] 
\; | N(p_1, \sigma_1) \rangle
\nonumber \\
&=& \bar u_2 \left[ \gamma^+ \; H_\pi^V (y, t) 
\; - \; \frac{E_\pi^V (y, t)}{2 M_N} \;
\sigma^{+\nu} \Delta_\nu \right] u_1 ,
\label{H_pi_E_pi_def}
\ee
where $\xi$ is the 4--vector of the space--time separation of the
fields and $\stackrel{\leftrightarrow}{\partial}{}^{\!\! \mu}
\equiv (\stackrel{\rightarrow}{\partial}{}^{\!\! \mu} 
- \stackrel{\leftarrow}{\partial}{}^{\!\! \mu})/2$.
Equation~(\ref{H_pi_E_pi_def}) applies in the parametric regime of 
pion momentum fractions $y = O(M_\pi / M_N)$ and momentum transfers 
$t = O(M_\pi^2)$ and is to be evaluated in chiral EFT. The pionic 
operator on the left--hand side has isovector quantum numbers, and the 
matrix element is understood in the sense of Eq.~(\ref{isospin}). 
The GPDs are denoted by $H_\pi^V$ and $E_\pi^V$ in accordance 
with the standard convention \cite{Ji:1996nm}, and are defined in
the interval $-1 < y < 1$. It is easy to see that the pionic light--ray 
operator is symmetric under $\xi \rightarrow -\xi$, whence 
the isovector GPDs are even functions of $y$,
\beq
H_\pi^V (y, t) \; = \; H_\pi^V (-y, t) , \hspace{2em}
E_\pi^V (y, t) \; = \; E_\pi^V (-y, t) .
\label{gpd_even}
\eeq
Note that for $\xi = 0$ the pionic operator reduces to the local
vector current of the pion field,
\beq
J^{+}(0) \;\; = \;\; \sum_{ab} \; \epsilon^{3ab} \; \pi^a (0) ,
\; \partial^+ \pi^b (0) ,
\eeq
the matrix element of which determines the chiral $\pi\pi$ cut 
contribution to the Dirac and Pauli form factors and was calculated
in Sec.~\ref{subsec:peripheral_processes}. When integrating 
Eq.~(\ref{H_pi_E_pi_def}) over $y$ the exponential factor produces
a delta function which enforces $\xi^- = 0$ and thus $\xi = 0$, so
that the matrix element of the light--ray operator
becomes that of the local vector current.
In this sense Eq.~(\ref{H_pi_E_pi_def}) is just a particular
representation, differential in the pion plus momentum fraction $y$,
of the chiral $\pi\pi$ cut in the current matrix element. 

The transverse coordinate--dependent (or impact parameter--dependent) 
distributions of soft pions in the nucleon are then defined in
analogy to the transverse charge and current densities, Eq.~(\ref{rho_def}),
as
\be
\left.
\begin{array}{l}
H_\pi^V (y, t = -\bm{\Delta}_T^2 ) 
\\[3ex] 
E_\pi^V (y, t = -\bm{\Delta}_T^2 )
\end{array}
\right\}
&=& \int d^2 b \; 
e^{i \bm{\Delta}_T \bm{b}}
\left\{
\begin{array}{r}
f_{1\pi}^V (y, b)
\\[3ex]
f_{2\pi}^V (y, b)
\end{array}
\right\} .
\label{f_pi_N_def}
\ee
It is convenient to introduce a modified helicity--flip distribution
in analogy to $\widetilde \rho_2$, Eq.~(\ref{rho_2_tilde_def}),
\be
\widetilde f_{2\pi}^V (y, b) &\equiv& \frac{\partial}{\partial b} 
\left[ \frac{f_{2\pi}^V (y, b)}{2 M_N} \right] .
\ee
The functions $f_{1\pi}^V (y, b)$ and $\widetilde f_{2\pi}^V (y, b)$ 
describe the isovector transverse spatial distribution of pions with 
plus momentum fraction $y$ and refer to the parametric regime 
$b = O(M_\pi^{-1})$ and $y = O(M_\pi/M_N)$. The interpretation of these 
spatial distributions in a transversely polarized nucleon is analogous 
to that of the transverse densities $\rho_1^V (b)$ and 
$\widetilde \rho_2^V (b)$,
cf.~Eq.~(\ref{j_plus_rho}) and Fig.~\ref{fig:interpretation}.

The peripheral pion distributions $f_{1\pi}^V (y, b)$ and 
$\widetilde f_{2\pi}^V (y, b)$ can be calculated in LO
chiral EFT in the same manner as the transverse densities
$\rho_1^V (b)$ and $\widetilde \rho_2^V (b)$, 
cf.~Sec.~\ref{sec:current_matrix_element}. One expresses the matrix 
element in Eq.~(\ref{H_pi_E_pi_def}) as a Feynman integral and separates 
it into the intermediate nucleon contribution and an effective contact 
term, cf.~Eqs.~(\ref{me_equivalent})--(\ref{me_contact}).
The intermediate nucleon contribution can then be represented as the
product of the light--front wave functions describing the transition 
of the initial and final nucleon to the $\pi N$ intermediate state.
For $y > 0$,
\be
\left. 
\begin{array}{l}
f_{1\pi}^V (y, b) \\[3ex] \widetilde f_{2\pi}^V (y, b) 
\end{array}
\right\}
&=& 
\frac{1}{2\pi y \bar y^3} 
\left\{
\begin{array}{c}
\displaystyle
[U_0(y,b/\bar{y})]^2 \; + \; 
[U_1(y,b/\bar{y})]^2
\\[3ex]
\displaystyle
-2 \, U_0(y,b/\bar{y})\; U_1 (y,b/\bar{y})
\end{array}
\right\} 
\hspace{2em} (y > 0),
\label{f_y_b_overlap}
\ee
where $U_{0, 1}$ are the coordinate--space wave functions defined 
in Eqs.~(\ref{psi_coordinate_decomposition}) and (\ref{psi_0_1_def});
for $y < 0$ one uses that [cf.~Eq.~(\ref{gpd_even})]
\beq
f_{1\pi}^V (y, b) \; = \; f_{1\pi}^V (-y, b),
\hspace{2em}
\widetilde f_{2\pi}^V (y, b) \; = \; \widetilde f_{2\pi}^V (-y, b).
\eeq
The contact term contribution is obtained from the Feynman integral as
\be
f_{1\pi}^V (y, b)_{\rm contact} &=& \delta (y) \; 
\rho_1^V (b)_{\rm contact} ,
\label{f_1pi_contact}
\ee
where $\rho_1^V (b)_{\rm contact}$ is given by Eq.~(\ref{rho_contact}).
This expression clearly identifies the contact term as a light--front 
``zero mode'' contribution. The complete isovector pion distribution 
$f_{1\pi}^V$ is then given by the sum of the overlap contribution 
Eq.~(\ref{f_y_b_overlap}) and the contact term Eq.~(\ref{f_1pi_contact}).
This representation of the chiral pion densities in particular
implies that [cf.\ Eqs.~(\ref{rho_overlap}) and
(\ref{rho_intermediate_contact})] 
\be
\left. 
\begin{array}{l}
\rho_1^V (b) \\[3ex] \widetilde\rho_2^V (b) 
\end{array}
\right\}
&=& 
\int_{-1}^1 dy
\left\{
\begin{array}{l}
f_{1\pi}^V (y, b) 
\\[3ex]
\widetilde f_{2\pi}^V (y, b)
\end{array} 
\right\} ,
\label{rho_overlap_f}
\ee
as is obvious from the above definition of $f_{1\pi}^V$ and $f_{2\pi}^V$.
In this sense our earlier results for the peripheral transverse densities
in chiral EFT have a straightforward interpretation as $y$--integrals 
of the peripheral pion GPDs in the nucleon. The significance of this 
connection lies in the fact that the peripheral pion GPDs have a more 
general physical significance and can in principle be measured 
independently in peripheral high--energy scattering processes.

\subsection{Charge density from peripheral partons}
We can now demonstrate the connection of the chiral component of the 
peripheral transverse densities with the peripheral quark/antiquark
content of the nucleon in QCD. Following Ref.~\cite{Strikman:2009bd}
the isovector quark/antiquark 
density in the nucleon at $b = O(M_\pi^{-1})$
generated by chiral dynamics is given by
\be
\left[u - d \right] (x, b)_{\rm chiral}
\;\; = \;\; 
\left[ \bar d - \bar u \right] (x, b)_{\rm chiral} 
&=& \int_x^1 \frac{dy}{y} \;
f_{1\pi}^V (y, b) \; q_\pi^{\text{val}} (z) 
\hspace{2em} (z  \equiv  x/y) , 
\label{conv_isovector}
\ee
where $z = x/y$ represents the fraction of the pion plus momentum
carried by the quark/antiquark, and $q_\pi^{\text{val}}(z)$ is the 
valence quark/antiquark density in the pion,
\be
q_\pi^{\text{val}} (z) &=& 
\pm \left[ \bar d - \bar u \right]_{\pi\pm} (z)
\;\; = \;\; \pm \left[ u - d \right]_{\pi\pm} (z) 
\;\; =\;\; \pm {\textstyle\frac{1}{2}} 
\left[ u - \bar u - d + \bar d \right]_{\pi\pm} (z) ,
\label{q_pi_val}
\ee
normalized such that
\be
\int_0^1 dz \, q_\pi^{\text{val}} (z) &=& 1 .
\label{valence_normalization}
\ee
Equation~(\ref{conv_isovector}) has the form of the usual partonic
convolution formulas and relies on the approximation that the
non-chiral transverse size of the pion can be neglected on
the scale $O(M_\pi^{-1})$; i.e., the spatial distribution of
peripheral quarks/antiquarks is determined entirely by the distribution 
of pions in the nucleon. The transverse charge density in the proton
is generally given by the integral of the quark minus antiquark densities 
over $x$, weighted by the quark charges ($e_u = 2/3, \, e_d = -1/3$),
\be
\rho_1^p(b) &=& \int_0^1 dx\; \left\{ e_u [u - \bar u](x, b) 
+ e_d [d - \bar d] (x, b) \right\} .
\label{rho_partonic_proton}
\ee
The isovector density is obtained by taking half the proton--neutron 
difference and using isospin symmetry
\be
\rho_1^V(b) &\equiv& {\textstyle\frac{1}{2}} [ \rho_1^p - \rho_1^n ](b)
\;\; = \;\; {\textstyle\frac{1}{2}}
(e_u - e_d) \int_0^1 dx\;  [u - \bar u - d + \bar d](x, b)
\hspace{2em} (e_u - e_d = 1)
\label{rho_partonic_isovector}
\ee
(this expression is valid even when including strange quarks in the 
individual proton and neutron densities). Substituting here the peripheral 
quark/antiquark densities generated by chiral dynamics, 
Eq.~(\ref{conv_isovector}), and using the normalization condition
Eq.~(\ref{valence_normalization}), one obtains
\be
\rho_1^V(b)_{\rm chiral} 
&=& \int_0^1 dx\;  [u - \bar u - d + \bar d](x, b)_{\rm chiral}
\nonumber \\
&=& \int_{0}^1 dx \, \int_x^1 \frac{dy}{y} \; f_{1\pi}^V (y, b) 
\; q_\pi^{\text{val}} (x/y) 
\nonumber \\
&=& 
\int_{0}^1 dy \, f_{1\pi}^V (y, b) \, \int_{0}^y \frac{dx}{y} 
\, q_\pi^{\text{val}} (x/y) 
\nonumber \\
&=& 
\int_{0}^1 dy \, f_{1\pi}^V (y, b) ,
\label{rho_1_partonic}
\ee
which agrees with Eq.~(\ref{rho_overlap_f}). It shows that resolving 
the peripheral pion into its quark/antiquark constituents and computing
the charge density from the quark/antiquark densities in the nucleon
leads to the same result as computing the charge density directly from the
distribution of (pointlike) pions --- as it should be in the parton picture.
The same applies, of course, to the spin--dependent density
$\widetilde\rho_2^V(b)$.

The $x \rightarrow 0$ limit in the integral Eq.~(\ref{rho_partonic_isovector})
and the role of the contact term in the charge density in the ``sum rule''
Eq.~(\ref{rho_1_partonic}) require special consideration. In the 
LO chiral expansion of the current matrix element and 
the correlator Eq.~(\ref{H_pi_E_pi_def}) it is supposed that the pion 
plus momentum fraction is of the parametric order $y = O(M_\pi/M_N)$. 
The contributions from $y \rightarrow 0$ on this scale are described 
by a delta function $\delta (y)$, i.e., one sees only their contribution 
to the total charge density ($y$--integral, first moment) but cannot 
resolve their dependence on $y$. In this sense the LO chiral 
expression for the peripheral parton densities is valid for $x = O(M_\pi/M_N)$
but otherwise not exceptionally small. Now calculating the charge density
requires integration down to $x \rightarrow 0$. It is clear from the
above that the LO chiral expansion and the limit 
$x \rightarrow 0$ do not commute. It explains why Eq.~(\ref{rho_1_partonic}) 
captures only the non-contact (intermediate $\pi N$) contribution to the
charge density, and why the contact term has to be added separately.
The LO chiral expansion correctly accounts for the total
charge density; it is just not smooth enough at $y \rightarrow 0$ to 
allow for the charge to be distributed over partons with finite $x$ 
and recovered by integration over $x$. A resummation of chiral EFT
at parametrically small $x$ in the logarithmic approximation has been
proposed in Refs.~\cite{Kivel:2007jj,Kivel:2008ry,Perevalova:2011qi};
this approach ``resolves'' the delta function of the finite--order
approximation into a finite--width function that allows integration 
over $x$. We emphasize that the small--$x$ problem discussed here is 
largely formal, and that chiral dynamics likely does not dominate
the actual small--$x$ behavior of peripheral parton densities
(for a discussion of chiral dynamics and resummation in the light 
of conventional small--$x$ physics, see Ref.~\cite{Strikman:2009bd}).
\section{Summary and outlook}
Using the light--front representation of relativistic dynamics we have 
expressed the LO chiral EFT results for the nucleon's 
peripheral transverse densities in ``first--quantized'' form, as overlap 
integrals of the light--front wave functions of a chiral $\pi N$ system.
The new representation is exactly equivalent to the ``second--quantized''
field--theoretical results and enables an intuitive understanding of
chiral dynamics in close analogy to non-relativistic quantum mechanics.
It reveals an inequality between the spin--independent and --dependent 
transverse densities, $|\widetilde\rho_2^V (b)| < \rho_1^V (b)$, which 
constrains the spatial distribution of charge and magnetization in
the nucleon. It also offers a simple dynamical explanation why the 
inequality is almost saturated. The wave function representation
permits straightforward numerical evaluation of the transverse 
densities and the underlying plus momentum distributions.
It also connects the peripheral transverse charge density with the
nucleon's peripheral partonic content (GPDs) generated by chiral dynamics. 

Our studies reveal two interesting general aspects of chiral dynamics 
and peripheral nucleon structure. One is the essentially 
relativistic character of chiral dynamics in the parametric
region of pion momenta $k = O(M_\pi)$. The large left--right asymmetry
of the chiral densities in the transversely polarized nucleon is a 
genuine relativistic effect and results in a ratio very different
from the non-relativistic estimate. It predicts the approximate 
equality $|\widetilde\rho_2^V (b)| \approx \rho_1^V(b)$ at 
$b = \textrm{few}\; M_\pi^{-1}$, which can be tested experimentally. 
In this sense measurements of the nucleon's Dirac and Pauli form factors 
can directly attest to the relativistic nature of chiral dynamics in 
the nucleon's periphery (for a discussion of the prospects for 
extracting the peripheral transverse densities from nucleon 
form factor data, see Refs.~\cite{Strikman:2010pu,Granados:2013moa}).

The other interesting aspect is the role of the pion's orbital angular 
momentum in peripheral nucleon structure. It is seen most clearly in 
the case of transverse nucleon polarization, where a single pion orbital 
with $L = 1$ accounts for both the spin--independent and --dependent 
densities and explains their properties.
This places chiral dynamics in the context of contemporary studies of 
orbital angular momentum in relativistic systems and quantum field theory,
inspired by the nucleon spin problem (for a recent review, see
Ref.~\cite{Leader:2013jra}). The first--quantized light--front 
representation is an essential tool in defining the angular momentum
content of relativistic systems and interpreting the dynamics,
and is therefore natural for chiral dynamics.

The chiral two--pion exchange contribution studied in this work affects 
the isovector component of the nucleon charge densities. In the isoscalar
component the chiral contribution starts with three--pion exchange and
is strongly suppressed, because of its shorter range and its higher 
order in the chiral expansion. In this sense the isovector component
computed here determines the large--distance behavior also of the 
individual proton and neutron densities, which are the sum of isovector
and isoscalar components. The analysis of experimental data for the
proton and neutron form factors at low $|t|$ should be done with  
dispersion--based parametrizations \cite{Belushkin:2006qa,Lorenz:2012tm}, 
which have correct analytic properties (singularity structure) and 
smoothly combine the two--pion exchange contribution with the vector 
meson resonances determining the bulk of the 
transverse densities \cite{Miller:2011du}.

The methods developed in the present work can be applied to several
related problems in nucleon structure. Of particular interest would be
the study of chiral dynamics in the matrix elements of the energy--momentum 
tensor, whose transverse densities describe the spatial distributions
of matter, momentum, and stress (or forces) in the nucleon.
This would in particular allow one to confront the ``particle--based'' 
definition of orbital angular momentum in the light--front representation
with the ``field--theoretical'' definition in terms of the energy--momentum
tensor, and in this way test the various angular momentum sum rules
proposed in the recent
literature \cite{Ji:1996ek,Ji:2012sj,Ji:2012ba,Ji:2012vj,Leader:2011cr}.

The light--front wave function representation of peripheral densities
could also be extended to $\Delta$ isobar intermediate states. The $\Delta$ 
contribution to the transverse densities was computed in the dispersive 
approach in Refs.~\cite{Strikman:2010pu,Granados:2013moa}. While is is 
numerically small at distances $b = \textrm{few}\, M_\pi^{-1}$ it plays
an important role in ensuring the proper scaling behavior of the
peripheral densities in the large--$N_c$ limit of QCD, where 
the $N$--$\Delta$ mass splitting scales as $M_\Delta - M_N = O(N_c^{-1})$.
The light--front wave functions for the $N \rightarrow \pi + \Delta$
transition can be defined in analogy to the $N \rightarrow \pi + N$ ones,
cf.\ Eq.~(\ref{psi_def}). Light--front time--ordered calculations 
with higher--spin particles are generally plagued by ultraviolet 
divergences resulting from the breaking of rotational invariance. 
An advantage of the Lorentz--invariant approach taken in the present
work (see Sec.~\ref{sec:current_matrix_element}), in which the light--front 
representation is derived from the Feynman integrals, is that 
it maintains rotational invariance and avoids such divergences.
It is thus particularly suited to including $\Delta$ intermediate states. 

The nucleon's chiral component can be probed also in high--energy 
scattering processes in $\gamma N, eN, \pi N$, or $NN$ scattering 
(squared center--of--mass energies $s \gg 1\, \textrm{GeV}^2$), 
by selecting reaction channels and 
kinematic regions where the process happens predominantly on a pion at 
transverse distances $b = O(M_\pi^{-1})$. Under certain conditions the 
amplitude for such processes can be expressed in terms of the
light--front wave functions of the peripheral $\pi N$ system and the 
amplitude for the high--energy scattering process on the pion.
These are the same light--front wave functions as those 
introduced in the present study of peripheral current matrix elements. 
Only the light--front formulation of chiral dynamics makes it possible 
to establish such a connection between low--energy and high--energy 
processes. This new connection greatly enlarges the number of experimental 
probes of chiral dynamics. 

One example is hard exclusive electroproduction of mesons on a 
peripheral pion, $e + N \rightarrow e' + \pi + \textrm{meson} + N'$
at $Q^2 \gg 1\, \textrm{GeV}^2$ and $x \ll M_\pi / M_N$, in the region 
where the invariant momentum transfer between the initial and final 
nucleons is $t_{NN'} \equiv (p_N - p_N')^2 = O(M_\pi^2)$. 
Here the high--energy process on 
the pion, $e + \pi \rightarrow e' + \textrm{meson} + \pi$, probes the 
quark/gluon GPDs in the pion \cite{Strikman:2003gz}. 
Such measurements could be performed
at a future Electron--Ion Collider (EIC) with appropriate forward 
detectors for the pion and the recoiling nucleon. Another example is 
wide--angle quasi--elastic scattering on a peripheral pion, 
$\pi + N \rightarrow \pi + \pi + N'$, 
in the region where $t_{NN'} = O(M_\pi^2)$;
here the high--energy process is described by the 
$\pi + \pi \rightarrow \pi + \pi$ elastic scattering amplitude
at squared center--of--mass energies $s_{\pi\pi} \gg 1\, \textrm{GeV}^2$. 
An interesting consequence of chiral dynamics is that the cross sections
for such ``pion exchange'' processes should have large transverse 
single--spin asymmetries, governed by the transverse spin dependence
of the chiral $\pi N$ light--front wave function. This circumstance 
might further help to distinguish such processes from conventional 
``vacuum exchange'' processes.
\section*{Acknowledgments}
Notice: Authored by Jefferson Science Associates, 
LLC under U.S.\ DOE Contract No.~DE-AC05-06OR23177. The U.S.\ Government 
retains a non--exclusive, paid--up, irrevocable, world--wide license to 
publish or reproduce this manuscript for U.S.\ Government purposes.
\appendix
\section{Light--front time--ordered formulation}
\label{app:time-ordered}
\noindent
In Sec.~\ref{sec:current_matrix_element} the light--front wave function 
of the chiral $\pi N$ system is introduced as an element of a particular
3--dimensional reduction of the Feynman integrals for the nucleon form 
factor in relativistically invariant chiral EFT. In this appendix we show 
that this object is identical to the conventional wave function, defined 
as a transition matrix element in light--front time--ordered perturbation 
theory \cite{Brodsky:1997de}. The correspondence is useful for relating 
our results to more 
phenomenological applications of light--front time--ordered perturbation 
theory, and for an eventual time--ordered formulation of chiral EFT.

In light--front quantization we consider the evolution of chiral effective 
field theory in light--front time $x^+ = t + z$. It is governed by the
Hamiltonian $H \equiv P^-/2$, where $P^- \equiv P^0 - P^z$ is the minus
component of the energy--momentum 4--vector of the field theory. 
In the interaction picture the Hamiltonian is split into a free part and 
an interaction, $H = H_0 + H_{\rm int}$. The eigenstates of the free 
Hamiltonian (which we denote as $|...\rangle_0$) are 
nucleon and pion single--particle states, characterized by their light--front
plus and transverse momenta, with eigenvalues given by the light--front
energies,
\be
H_0 \, | N(p, \sigma) \rangle_0 &=& \frac{p^-}{2} \, | N(p, \sigma) \rangle_0,
\hspace{2em} p^- \;\; \equiv \;\; 
\frac{\bm{p}_T^2 + M_N^2}{2 p^+} ,
\\[1ex]
H_0 \, | \pi(k) \rangle_0 &=& \frac{k^-}{2} \, | \pi(k) \rangle_0,
\hspace{2em} k^- \;\; \equiv \;\; 
\frac{\bm{k}_T^2 + M_\pi^2}{2 k^+} ,
\ee
and are normalized as
\be
{}_0\langle N (p', \sigma')| N(p, \sigma) \rangle_0
&=& 2 p^+ \; (2\pi)^3 \; \delta (p^{\prime +} - p^+) \;
\delta^{(2)} (\bm{p}_{T}' - \bm{p}_{T}) \;
\delta (\sigma', \sigma) ,
\\
{}_0\langle \pi (k')| \pi(k) \rangle_0
&=& 2 k^+ \; (2\pi)^3 \; \delta (k^{\prime +} - k^+) \;
\delta^{(2)} (\bm{k}_{T}' - \bm{k}_{T}) .
\ee
The interactions represent an operator in the product space spanned 
by the single--particle states and induce transitions between product
states with different particle number. The evolution of a state at 
light--front time $x_1^+$ to $x_2^+$ is described by the action of the 
time evolution operator
\be
S(x_2^+, x_1^+) &\equiv& {\rm T} \; \exp\left[ 
-i \int_{x_1^+}^{x_2^+} dx^+ H_{\rm int}(x^+) \right],
\\[1ex]
H_{\rm int}(x^+) &\equiv& \exp(i H_0 x^+) \; H_{\rm int} \; \exp(-i H_0 x^+) ,
\ee
where the time ordering is in $x^+$ and it is assumed that the original 
interaction operator $H_{\rm int}$ does not explicitly depend on $x^+$.

Consider now the matrix element of the current operator at time $x^+ = 0$
between physical nucleon states with plus momenta $p_1^+ = p_2^+ = p^+$ 
and transverse momenta such that $\bm{p}_{2T} - \bm{p}_{1T} = \bm{\Delta}_T$, 
cf.\ Eq.~(\ref{transverse_frame}). Assuming that the interaction is
switched off adiabatically at $x^+ \rightarrow \pm\infty$, the
matrix element is given by 
\be
\langle N_2 | \, J^+(0) \, | N_1 \rangle
&\equiv& \langle N (p_2, \sigma_2 ) | \, J^+(0) \, 
| N (p_1, \sigma_1 ) \rangle
\nonumber
\\[1ex]
&=& {}_0\langle N (p_2, \sigma_2 ) | \, S(\infty, 0) \, J^+(0) \,
S(0, -\infty) \, | N (p_1, \sigma_1 ) \rangle_0 .
\label{me_topt}
\ee
A peripheral contribution originates from the components of the evolved
nucleon state that contain a peripheral pion. In leading order of the 
interaction $H_{\rm int}$ this component is a state with a nucleon and a 
single peripheral pion. The corresponding contribution to the matrix element 
is obtained by inserting free pion--nucleon intermediate
states to the left and right of the current operator,
\be
\langle N_2 | \, J^+(0) \, | N_1 \rangle
&=& 
\int \frac{dk_1^+}{(2\pi) 2 k_1^+}
\int \frac{d^2 k_{1T}}{(2\pi)^2}
\;
\int \frac{dk_2^+}{(2\pi) 2 k_2^+}
\int \frac{d^2 k_{2T}}{(2\pi)^2}
\;
\int \frac{dl^+}{(2\pi) 2 l^+}
\int \frac{d^2 l_{T}}{(2\pi)^2}
\nonumber 
\\[1ex]
&\times& 
{}_0\langle N (p_2, \sigma_2 ) | \, S(\infty, 0) \, 
| \pi (k_2 ) N(l, \sigma) \rangle_0
\nonumber 
\\[1ex]
&\times & {}_0\langle \pi (k_2 ) | \, J^+(0) \, | \pi (k_1 ) \rangle_0
\nonumber
\\[1ex]
&\times& 
{}_0\langle \pi (k_1 ) N(l, \sigma) | \,
S(0, -\infty) \, | N (p_1, \sigma_1 ) \rangle_0 .
\label{me_topt_intermediate}
\ee
The amplitude for the transition from the initial free nucleon state
to the pion--nucleon intermediate state at time $x^+ = 0$ is
\be
\lefteqn{
{}_0\langle \pi (k_1 ) N(l, \sigma) | \, S(0, -\infty)_{\rm LO} \, 
| N(p_1, \sigma_1) \rangle_0}
\nonumber
\\[1ex]
&=& {}_0\langle \pi (k_1) N(l, \sigma) | \, (-i) \int_{-\infty}^{\, 0} 
dx^+ H_{\rm int}(x^+) \, | N(p_1, \sigma_1) \rangle_0
\nonumber
\\[1ex]
&=& {}_0\langle \pi (k_1) N(l, \sigma) | \, H_{\rm int} \, 
| N(p_1, \sigma_1) \rangle_0 \;
(-i) \int_{-\infty}^{\, 0} 
dx^+ \; e^{i (k_1^- + l^- - p_1^- - i0) x^+ /2} 
\nonumber
\\[1ex]
&=& - \, \frac{{}_0\langle \pi (k_1) N(l, \sigma) | \, H_{\rm int} 
\, | N(p_1, \sigma_1) 
\rangle_0}
{\frac{1}{2}(k_1^- + l^- - p_1^-)}
\nonumber
\\[1ex]
&& \left( p_1^- \; = \; \frac{\bm{p}_{1T}^2 + M_N^2}{p_1^+} ,
\;\;
k_1^- \; = \; \frac{\bm{k}_{1T}^2 + M_\pi^2}{k_1^+} ,
\;\;
l^- \; = \; \frac{\bm{l}_{T}^2 + M_N^2}{l^+} \right) ,
\label{transition_me}
\ee
where the infinitesimal imaginary part of the light--front energies
implies that the interaction is switched on adiabatically
as $x^+$ increases from $-\infty$. The interaction Hamiltonian is the 
integral of the interaction Hamiltonian density, which in turn is
the negative of the interaction Lagrangian density,
\be
H_{\rm int} 
&=& \frac{1}{2} \int dx^- d^2 x_T \; \mathcal{H}_{\rm int}(x)_{x^+ = 0}
\;\; = \;\; - \frac{1}{2}
\int dx^- d^2 x_T \; \mathcal{L}_{\rm int}(x)_{x^+ = 0} .
\ee
We use translational invariance to evaluate the matrix element
in Eq.~(\ref{transition_me}),
\be
\lefteqn{{}_0\langle \pi (k_1) N(l, \sigma) | \, H_{\rm int} \,
| N(p_1, \sigma_1) 
\rangle_0}
\nonumber
\\[1ex]
 &=& 
- \frac{1}{2}
\int dx^- d^2 x_T \; 
{}_0\langle \pi (k_1) N(l, \sigma) | \, \mathcal{L}_{\rm int}(x) \,
| N(p_1, \sigma_1) \rangle_0
\nonumber
\\
&=& - \frac{1}{2}
\int dx^- d^2 x_T \; e^{i(k_1 + l - p_1)x} \;
{}_0\langle \pi (k_1) N(l, \sigma) | \, \mathcal{L}_{\rm int}(0) \,
| N(p_1, \sigma_1) \rangle_0
\nonumber 
\\[1ex]
&=& - (2\pi^3) \, \delta(k_1^+ + l^+ - p_1^+) \;
\delta^{(2)}(\bm{k}_{1T} + \bm{l}_T - \bm{p}_{1T}) \;
{}_0 \langle \pi (k_1) N(l, \sigma) | \, \mathcal{L}_{\rm int}(0) \,
| N(p_1, \sigma_1) \rangle_0 .
\ee
The LO transition matrix element Eq.~(\ref{transition_me}) 
can thus be represented as
\be
{}_0\langle \pi (k_1 ) N(l, \sigma) | \,
S(0, -\infty)_{\rm LO} \, | N(p_1, \sigma_1) \rangle_0
&=&  2 p_1^+ \; (2\pi^3) \, \delta(k_1^+ + l^+ - p_1^+) \;
\delta^{(2)}(\bm{k}_{1T} + \bm{l}_T - \bm{p}_{1T}) \; \Psi ,
\label{transition_wave_function}
\\[2ex]
\Psi \;\; \equiv \;\; \Psi (y, \bm{k}_T, \bm{p}_{1T}; \sigma, \sigma_1) 
& \equiv &
\frac{{}_0\langle \pi (k_1) N(l, \sigma) | \, \mathcal{L}_{\rm int}(0) \,
| N(p_1, \sigma_1) \rangle_0}{p_1^+ (k_1^- + l^- - p_1^-)} ,
\label{wave_function_topt}
\ee
where $\Psi$ is the light--front wave function of the $\pi N$ component of 
the nucleon. It is invariant under longitudinal boosts and can be regarded 
as a function of the independent momentum variables $y = k_1^+/p_1^+, 
\bm{k}_T = \bm{k}_{1T} - \bm{p}_{1T} = -\bm{l}_T$, and $\bm{p}_{1T}$. 
An analogous expression can be written for the complex conjugate
matrix element describing the transition from the outgoing free
pion--nucleon intermediate state to the final free nucleon state.

The wave function defined by 
Eq.~(\ref{wave_function_topt}) is identical to the one 
introduced in the context of the reduction of the Feynman 
integral, Eq.~(\ref{psi_def}). Namely, the light--front energy denominator 
in Eq.~(\ref{wave_function_topt}) is just the invariant mass 
difference Eq.~(\ref{invariant_mass_explicit}),
\be
p_1^+ (k_1^- + l^- - p_1^-) &=& 
\frac{\bm{k}_{1T}^2 + M_\pi^2}{y} + 
\frac{\bm{l}_{T}^2 + M_N^2}{1 - y} - M_N^2 - \bm{p}_{1T}^2 
\;\; = \;\;
\Delta \mathcal{M}^2.
\ee
Furthermore, the transition matrix element of the Lagrangian density 
in the numerator in Eq.~(\ref{wave_function_topt}) is just the 
LO $\pi N$ vertex function Eq.~(\ref{Gamma}),
\be
{}_0\langle \pi (k_1) N(l, \sigma) | \, \mathcal{L}_{\rm int}(0) \,
| N(p_1, \sigma_1) \rangle_0
&=& \Gamma (y, \bm{k}_T, \bm{p}_{1T}; \sigma, \sigma_{1})
\;\; + \;\; 
\textrm{terms} \propto (k_1^- + l^- - p_1^-) .
\label{me_correspondence}
\ee
The expression quoted here is for the $p \rightarrow \pi^0 + p$ isospin 
component of the matrix element; the other components follow from isospin 
invariance [cf.\ the comments after Eq.~(\ref{me_onshell_overlap})].

The final result for the leading--order peripheral contribution 
to the current matrix in the time--ordered approach is then obtained 
by (a)~substituting the transition matrix elements in 
Eq.~(\ref{me_topt_intermediate}) by the
representation Eq.~(\ref{transition_wave_function}), with the isospin
factor $\sqrt{2}$ for the transitions $p \rightarrow \pi^+ + n$ and
$n \rightarrow \pi^- + p$; (b)~substituting the explicit expression for
the matrix element of the current operator between free (pointlike) 
charged pion states,
${}_0\langle\pi^{(\pm)}(k_2) | \, J^+(0) \, | \pi^{(\pm)} (k_1) \rangle_{0} 
= \pm (k_1^+ + k_2^+)$; (c)~integrating 
over the redundant intermediate pion and nucleon momenta using the 
delta functions in Eq.~(\ref{transition_wave_function}). The result for the
isovector component is identical to Eq.~(\ref{me_onshell_overlap}).
Altogether this establishes the correspondence between the time--ordered 
and the invariant calculation of the LO chiral contribution to the
peripheral densities.

Note that the transition matrix element Eq.~(\ref{me_correspondence}) 
calculated with the axial vector $\pi NN$ coupling 
(as appears in the original chiral Lagrangian) and 
the pseudoscalar coupling (as emerges from our reduction of the 
Feynman integral) differ by a term 
$\propto (k_1^- + l^- - p_1^-)$. This term cancels the energy 
denominator of the wave function and therefore results in a
contact term in the matrix element, in agreement with the findings of 
Sec.~\ref{sec:current_matrix_element}. One thus understands 
how the effective contact term Eq.~(\ref{me_contact}) appears in the 
time--ordered formulation. Calculation of the complete contact term 
contribution in the time--ordered formulation would be possible with
a careful limiting procedure for the instantaneous exchanges.
An advantage of our invariant formulation is that it allows us to
calculate this contribution with minimum effort.
\end{document}